\documentstyle[aps,psfig]{revtex}
\makeatletter
\newif\ifpr@pstyle \pr@pstylefalse
\newif\ifnons@qeq  \nons@qeqfalse
\def\bigpage{
	\newlength{\paperbaselineskip}
	\setlength{\paperbaselineskip}{20pt plus 0.2pt minus 0.1pt}
	\def\@oddfoot{\hfil\footnotesize -- \thepage~--\hfil}
	\let\@evenfoot\@oddfoot
        \def\thesection{\Roman{section}}
        \def\thesubsection{\Alph{subsection}}
        \def\@ourappendix{\par\setcounter{section}{0}
                      \setcounter{subsection}{0}
                      \def\thesection{\Alph{section}}
                      \ifnons@qeq
                      \def\theequation{\Alph{section}.\arabic{equation}}\fi}
        \def\appendix{\@ourappendix}
        \def\section{\@startsection {section}{1}%
            {\z@}{5ex plus .2ex minus .4ex}%
            {1.5ex plus.4ex minus .1ex}%
            {\centering\ifpr@pstyle\else\ifx\undefined\reset@font\else%
             \reset@font\fi\large\fi\bf}}
        \def\subsection{\@startsection{subsection}%
            {2}{\z@}{3.25ex plus .4ex minus .4ex}%
            {1ex plus .2ex}
            {\centering\ifpr@pstyle\else\ifx\undefined\reset@font\else%
             \reset@font\fi\normalsize\fi\bf}}
}
\bigpage
\newfont{\fourteencp}{cmcsc10 scaled\magstep2}
\newfont{\titlefont}{cmbx10 scaled\magstep2}
\newfont{\authorfont}{cmcsc10 scaled\magstep1}
\newfont{\fourteenmib}{cmmib10 scaled\magstep2}
	\skewchar\fourteenmib='177
\newfont{\elevenmib}{cmmib10 scaled\magstephalf}
	\skewchar\elevenmib='177
\newfont{\ninemib}{cmmib9} \skewchar\ninemib='177
\makeatletter
\newcommand\nonsequentialeqnum{
        \nons@qeqtrue
	\@addtoreset{equation}{section}
	\def\theequation{\arabic{section}.\arabic{equation}}}
\newif\ifp@bblock  \p@bblocktrue
\newcommand\nopubblock{\p@bblockfalse}
\newcommand\topspace{\hrule height 0pt depth 0pt \vskip}
\newcommand\p@bblock{\begingroup \tabskip=\hsize minus \hsize
	\baselineskip=1.5\ht\strutbox \topspace-2\baselineskip
	\halign to\hsize{\strut ##\hfil\tabskip=0pt\crcr
	\the\Pubnum\crcr\the\date\crcr}\endgroup}
\newcommand\YUKAWAmark{\hbox{
        \ifpr@pstyle\ninemib\else\elevenmib\fi
        Yukawa\hskip1mm Institute\hskip1mm Kyoto \hfill}}
\newtoks\date
\newtoks\Pubnum
\let\pubnum=\Pubnum
\Pubnum={}
\date={\today}
\newcommand{\frontpageskip}{\vspace{12pt plus .5fil minus 2pt}}
\def\@authoraddress{} \def\@title{}
\def\title#1{\gdef\@title{\frontpageskip
	\begin{center}{\titlefont #1}\end{center}\par}}
\def\@author#1{\frontpageskip\par\begin{center}{\authorfont #1}
	\end{center}
	\nobreak}
\def\author#1{\expandafter\def\expandafter\@authoraddress\expandafter
    {\@authoraddress{\@author{#1}}}}
\def\andauthor#1{\expandafter\def\expandafter\@authoraddress\expandafter
    {\@authoraddress{\frontpageskip\centerline{and}\@author{#1}}}}
\def\authors#1{\expandafter\def\expandafter\@authoraddress\expandafter
    {\@authoraddress{\frontpageskip\noindent #1}}}
\def\@address#1{\par\begin{center}{\sl #1}\end{center}\par}
\def\address#1{\expandafter\def\expandafter\@authoraddress\expandafter
    {\@authoraddress{\@address{#1}}}}
\def\andaddress#1{\expandafter\def\expandafter%
    \@authoraddress\expandafter
    {\@authoraddress{\par\centerline{\sl and}\@address{#1}}}}
\renewcommand{\thanks}[1]{\footnote{#1}}
\def\maketitle{\par
  \begingroup
       \def\thefootnote{\fnsymbol{footnote}}
	\thispagestyle{empty}
        \baselineskip=\paperbaselineskip
	\@maketitle
	\endgroup
	\setcounter{footnote}{0}
	\let\maketitle\relax \let\@maketitle\relax
	\let\@thanks\relax \let\@title\relax
	\let\@title\relax \let\@authoraddress\relax
	\let\thanks\relax}
\def\@maketitle{%
        \ifpr@pstyle\vspace{-1.0cm}\else\vspace{-1.7cm}\fi
	\YUKAWAmark\vskip0.6cm
	\ifp@bblock\p@bblock \else\hrule height 0pt \relax \fi
	\@title
	\@authoraddress
	}
\makeatother
\begin{document}

\pubnum{YITP-99-13}
\date{February 1999}
\title{Scaling Analysis of Fluctuating Strength Function}
\author{Hirokazu Aiba}
\address{Koka Women's College, 38 Kadono-cho Nishikyogoku, Ukyo-ku,
615 Kyoto, Japan}
\andauthor{Masayuki Matsuo}
\address{Yukawa Institute for Theoretical Physics, Kyoto University,
601-01 Kyoto, Japan}
\maketitle
\vfill
%\renewcommand{\abstract}{\par\frontpageskip\centerline{
%             \ifpr@pstyle\twelvecp\else\fourteencp\fi Abstract}
%	\vspace{8pt plus 3pt minus 3pt}}
%\begin{abstract}
\centerline{\fourteencp Abstract}
\vspace{8pt}
We propose a new method 
to analyze fluctuations in the strength function phenomena in
highly excited nuclei.  
Extending the method of multifractal analysis to the cases 
where the strength fluctuations do not obey power scaling laws,
we introduce a new measure of fluctuation, 
called the local scaling dimension,  which characterizes
scaling behavior of the strength fluctuation as a function
of energy bin width subdividing the strength function.
We discuss properties of the new measure by  applying it to
a model system which simulates the doorway damping mechanism 
of giant resonances. It is found that the local scaling dimension
characterizes well fluctuations and their energy scales of
fine structures in the strength function associated with 
the damped collective motions.
\vspace{30mm}
%\end{abstract}
%\pacs{PACS: 21.10.-k, 24.30.Cz, 24.60.-k, 24.60.Ky}

\section{Introduction}
\label{sec:intro}

Collective or single-particle modes of excitation 
which have high excitation energies usually
become damped motions due to  coupling with background compound states
with high level density. 
Strengths of exciting such modes form a broad peak (when plotted
as a function of the excitation energy of a nucleus) which are spread over 
an interval of excitation energy.
A typical example is the giant resonances which have a damping width
of MeV order while the resonance energy is several to a few tens MeV
depending on quantum numbers and nuclear masses \cite{bertsch,woude}.
Strength distribution plotted as a function of the excitation energy 
is called strength function.
The strength function usually shows 
a smooth profile such as Lorentzian shape, but it also exhibits 
fluctuations called fine structures 
when the strength function is studied in high resolution
experiments \cite{woude,hansen,kuhner,kilgus,kamer}.

An important origin of the damping of the giant resonances in heavy
nuclei is the doorway coupling \cite{bertsch}.
The collectivity of the excitation mode
is described well in terms of a coherent superposition
of 1p-1h excitations, as often prescribed by the RPA models.
On the other hand, the damping of the mode is caused by (besides 
the particle escaping whose effect is small in heavy nuclei) 
coupling to other kinds of excitation modes 
such as 2p-2h excitations or 1p-1h plus low-lying
vibrational modes, which are
called doorway states.
Although the doorway states couple further with more complicated states,
e.g. many-particle many-hole states and have their own spreading
width,  the total damping width of the giant resonances
is explained reasonably well by the doorway coupling
\cite{bertsch}. It is argued that the doorway states influence also 
the fine structure of the strength distribution\cite{kilgus,kamer}.
However, relation between the doorway states and the fine structures 
remains to be understood since it would depend on
the spreading of doorway states into more complicated states, 
which is not very well known.

If we look at the strength function with the finest energy scale 
dissolving individual energy eigenstates by assuming 
small particle escaping,  another mechanism of the strength
fluctuation is expected to emerge. 
Because of the chaotic nature of the compound
states at the high excitation energy, 
strengths associated with individual eigenstates would show
statistical fluctuation following the Porter-Thomas distribution
\cite{porter,brody}.
When the particle escaping width
is larger than mean level spacing so that individual level
peaks  overlap, the Ericson fluctuation
emerges\ \cite{brody,ericson,bohigas-weidenmuller}. 
However it is natural to expect that
these statistical behaviors based on the 
random matrix theories \cite{brody,dyson,mehta,bohigas} govern
the strength fluctuation only at small energy scales.
One should rather expect that the strength function
associated with the damping of giant resonances involves
different kinds of fluctuations coexisting at different energy scales.
(Note also there exists an argument that 
experimental fine 
structures of particle decay strength distribution
under a certain situation resembles large fluctuations in 
simulated spectra generated solely on the basis of the statistical 
Porter-Thomas fluctuation, known as the pandemonium
\cite{hansen,pandemonium}.
Although this is a another possible origin of the fine
structures, the property of the statistical fluctuation should 
depend on the level density
of compound states under discussion).
Thus it is important to 
analyze the fluctuations of strength function in connection with
the energy scales of the fluctuation.

In this paper, we propose an new method to perform such an fluctuation
analysis of the strength functions. 
With this method we characterize the strength fluctuation
with focus on its scaling behavior by taking a similar  
approach  widely adopted to analyze fluctuation
phenomena having self-similar multifractal
structure\cite{maccauly1,maccauly2,feder}.
The multifractal fluctuations accompany a characteristic scaling property
obeying the power laws, and they are found 
in various kinds of chaotic classical dynamical systems 
\cite{maccauly1,maccauly2,feder}.
Furthermore multifractal fluctuations in quantum 
wave functions are found recently for electrons in 
disordered matter\cite{aoki,castellani,janssen}. 
The idea is also applied to the strength function fluctuations 
\cite{aiba,gorski},
and an approximate power scaling is found in a shell model calculation
of the giant dipole resonances \cite{gorski}. 
In contrast to these multifractal analyses, however, we consider in
this paper more general situations 
where the fluctuation does not necessarily follow
the scaling laws, and we rather try to characterize the 
scaling behavior  which may be dependent on the energy scales
due to different coexisting fluctuation mechanisms.
For this purpose, we introduce a new measure of fluctuation
by extending the generalized fractal dimension 
\cite{hentschel-procaccia,halsey}
used in the standard multifractal analysis. The new measure, which we
call the local scaling dimension, characterizes the scaling property
as a function of the energy scale of fluctuation.
With use of simple examples, we will demonstrate in this paper 
how the local scaling dimension carries information on
the dynamics of the damping process.

For the analysis of the strength fluctuations, the autocorrelation
function has been used widely as a  measure of fluctuation
in various contexts. 
It was applied for example to analyze the Ericson fluctuation
in compound nuclear reactions to derive the particle decay 
width\ \cite{brody,ericson}, and also to the strength of giant resonances
and the decay particle spectra \cite{hansen,kilgus}.
The autocorrelation function and its Fourier transform is also used
to characterize energy level fluctuations in the random matrix
theories and chaotic quantal
systems \cite{leviander,wilkie,lombardi,alhassid}.  We will show that 
the present approach and the autocorrelation analysis are
related to each other and that the local scaling analysis can be used 
to reveal the energy scales involved in the strength fluctuation.

The present method treats in a single framework strength fluctuations
in wide range of the energy scale, covering the statistical
fluctuations at the finest scale around the level spacing order as
well as those associated with the doorway states at the larger scale.
In the present paper, therefore, we use the word fine structure to
denote all kinds of strength fluctuations including the doorway
fluctuations, which are sometimes called the intermediate
structures in the literature.

The paper is organized as follows.
In section\ \ref{sec:ldim}, we describe the formalism of
the fluctuation analysis and introduce the local scaling dimension.
In addition, relation to the autocorrelation function is clarified.
In section\ \ref{sec:numerical}, employing a schematic 
model which simulates the doorway damping mechanism of 
giant resonance, we 
discuss in detail how the local scaling dimension characterizes
fluctuations in the model strength function. In particular,
an emphasis is put on the relation between the profile of the local scaling
dimension and the energy scales involved in the doorway damping process.
Finally, section\ \ref{sec:conclusion} is devoted to conclusions.

\section{Scaling Analysis of Strength Function}
\label{sec:ldim}

\subsection{Local Scaling Dimension}
\label{sec:localscaling}

The strength function is expressed as \cite{Bohr-Mottelson}
\begin{equation}
S(E)=\sum_i S_i\delta(E-E_i+E_0),
\label{defstr}
\end{equation}
for exciting the nucleus with excitation energy $E$ 
by a probe which excite a mode under consideration 
if we neglect the coupling to continuum states of escaping particles
and experimental resolutions for simplicity.  
Here $E_i$ and $E_0$ are the energy of discrete
levels and the ground state energy of the whole nuclear
many-body Hamiltonian, respectively,
and $S_i$ denotes the strength of exciting the $i$-th energy level.
Let us assume strengths are normalized as $\sum_i S_i=1$.

By smoothing the strength function with respect to the excitation
 energy $E$ with a sufficiently large smoothing energy width, or 
by putting the strengths into energy bins with large bin width, 
a smooth profile (e.g. a Lorentzian shape) will show up. If small
smoothing width or bin width is used, resultant distribution will
exhibit fluctuation associated with fine structures of the original
strength function. Thus the fluctuation of the strength function
generally depends on the energy scale with which one measures. 
To evaluate the scale dependent fluctuation, we apply the
method described below. It is an extension of the 
scaling analysis widely used to characterize  multifractal structures\ 
\cite{maccauly1,maccauly2,feder}.

Let us consider binned distribution of the strength function $S(E)$ by
dividing whole energy interval under consideration 
into $L$ bins with length $\epsilon$. Strength
contained in $n$--th bin is denoted by $p_n$,
\begin{equation}
p_n\equiv\sum_{i\in n{\rm -th~ bin}}S_i.
\label{defp}
\end{equation}
To characterize the fluctuation of the binned strength 
distribution  $\{p_n\}$, 
moments of the distribution are introduced.
This leads to the so-called partition function
$\chi_m(\epsilon)$ defined by
\begin{equation}
\chi_m(\epsilon)\equiv \sum_{n=1}^L p_n^m \\
                 =L\langle p_n^m\rangle.
\label{partition}
\end{equation}
For completeness of the definition, 
we need to introduce an average of $\chi_m(\epsilon)$
with respect to the bin boundary. This is because
there is still arbitrariness in Eq.(\ref{partition}) concerning how
the boundaries of bins are chosen even if the bin width $\epsilon$
is fixed. To overcome this uncertainty, we make an average over
different choices of the bin boundaries. 
When the average is done
with sufficiently large numbers of choices so that the boundaries
cover densely the energy axis, the averaged partition function
becomes independent of the boundaries.

It is of special interest if the partition function has a 
scaling property. If we consider 
a trivial case where the strength is uniformly distributed
with no fluctuation, i.e.,
$p_n=const=1/L\propto \epsilon$ for all $L$ and $\epsilon$, then 
an power low scaling $\chi_m(\epsilon)\propto\epsilon^{(m-1)}$ 
holds for $\chi_m(\epsilon)$ with respect to the bin width $\epsilon$.
When the fluctuation is present, the partition function generally 
deviates from this limit. An extreme case of  large fluctuation
is the situation where the strength is
concentrated in a single energy level. In this case 
the partition function becomes constant $\chi_m(\epsilon)=1$ for any 
value of $\epsilon$ since the binned strength is then
$p_n=1$ for only one bin, and $p_{n'}=0$ for the others. In other words
the partition function $\chi_m(\epsilon)$ scales with zero power.
If the fluctuation shows a fractal structure, i.e.,
it has a self-similar structure against the change of
the scale,
the partition function shows a power scaling with power
different from the trivial cases, like
\begin{equation}
\chi_m(\epsilon)\propto\epsilon^{D_m^{\rm fractal}(m-1)},
\label{chifrac}
\end{equation}
or equivalently a linear scaling between $\log\chi_m(\epsilon)$
and $\log\epsilon$,
\begin{equation}
\log\chi_m(\epsilon)\approx (m-1){D_m^{\rm fractal}\log\epsilon}.
\label{chifraclog}
\end{equation}
The scaling coefficient $D_m^{\rm fractal}$ called the
generalized fractal dimension characterizes
the multifractal\ \cite{hentschel-procaccia,halsey}.
The trivial case of uniform distribution 
corresponds to $D_m^{\rm fractal}=1$, implying a one dimensional
uniform object in the $E$ axis. 
The case of the single sharp peak corresponds to
$D_m^{\rm fractal}=0$ since it is just a dot (zero dimensional
object).

It should be noted however that the partition function does not
follow the power scaling law in general. The fluctuation of the
binned strength may change when we change the bin width.
This will happen when the strength function contains fluctuations 
having specific energy scales.
For the case of the damping of giant resonance,  fluctuation may
reflect the doorway states such as 2p-2h 
configurations, for which their spreading width could be one of
the energy scales involved. Also, when we look at very
small energy scale comparable with average level spacing, 
strength fluctuation associated with individual
energy levels may emerge. Therefore, it is more useful to analyze 
``how the partition function scales for different energy scales''
 than to seek the power scaling property. For this purpose, we
introduce an extension of the generalized fractal dimension  by
defining the scaling coefficients as local linear coefficients
of $\log\chi_m(\epsilon)$ vs. $\log\epsilon$  at each energy
scale $\epsilon$, or equivalently defining by
\begin{equation}
D_m(\epsilon)\equiv {1\over m-1}
\frac{\partial\log\chi_m(\epsilon)}
{\partial\log\epsilon}.
\label{scaledim}
\end{equation}
We call this quantity the local scaling
dimension since it is defined at each energy scale $\epsilon$ and 
this is  a function of $\epsilon$.
Note also that if the local scaling 
dimension $D_m(\epsilon)$ is constant over
a long interval of the energy scale $\epsilon$, it implies presence
of the multifractal structure of the strength fluctuation.
In this case the local scaling
dimension reduces 
to the standard generalized fractal dimension $D_m^{\rm fractal}$ defined 
by Eq.\ (\ref{chifraclog}).

In actual calculation of the local scaling dimension, we 
define it by means of finite difference under the change
of a factor 2,
\begin{equation}
D_m(\sqrt{2}\epsilon) = {1\over m-1}\frac{
\log\chi_m(2\epsilon)-
\log\chi_m(\epsilon)}
{\log 2},
\label{approscaledim}
\end{equation}
rather than the derivative in Eq.\ (\ref{scaledim}).
Using the finite difference, the calculation is very simple for 
all the moments. 
The local scaling dimension $D_2(\epsilon)$ defined by the derivative
has an exact form (\ref{d2explicit}) expressed 
in terms of the individual 
energy levels $E_i$ and associated strengths ${S_i}$, as given in 
Appendix\ \ref{appa}.  However, 
exact calculation with use of
Eqs. (\ref{scaledim}) and (\ref{inichitheta2}) 
is more complicated than the calculation using
the finite difference\ (\ref{approscaledim}) except for the 
second moment local scaling dimension $D_2(\epsilon)$. 
Note also that the finite difference gives an effective smoothing.
We will discuss difference between
the two definitions in Sec.\ \ref{sec:results}.

\subsection{Schematic Examples}
\label{sec:exampes}
Let us illustrate how the local scaling dimension carries information 
on the fluctuation of strength function 
by using some schematic examples.

\subsubsection{Lorentzian}
\label{sec:lorentzian}

\begin{figure}
\begin{center}
  \begin{minipage}{7.5cm}
       \psfig{file=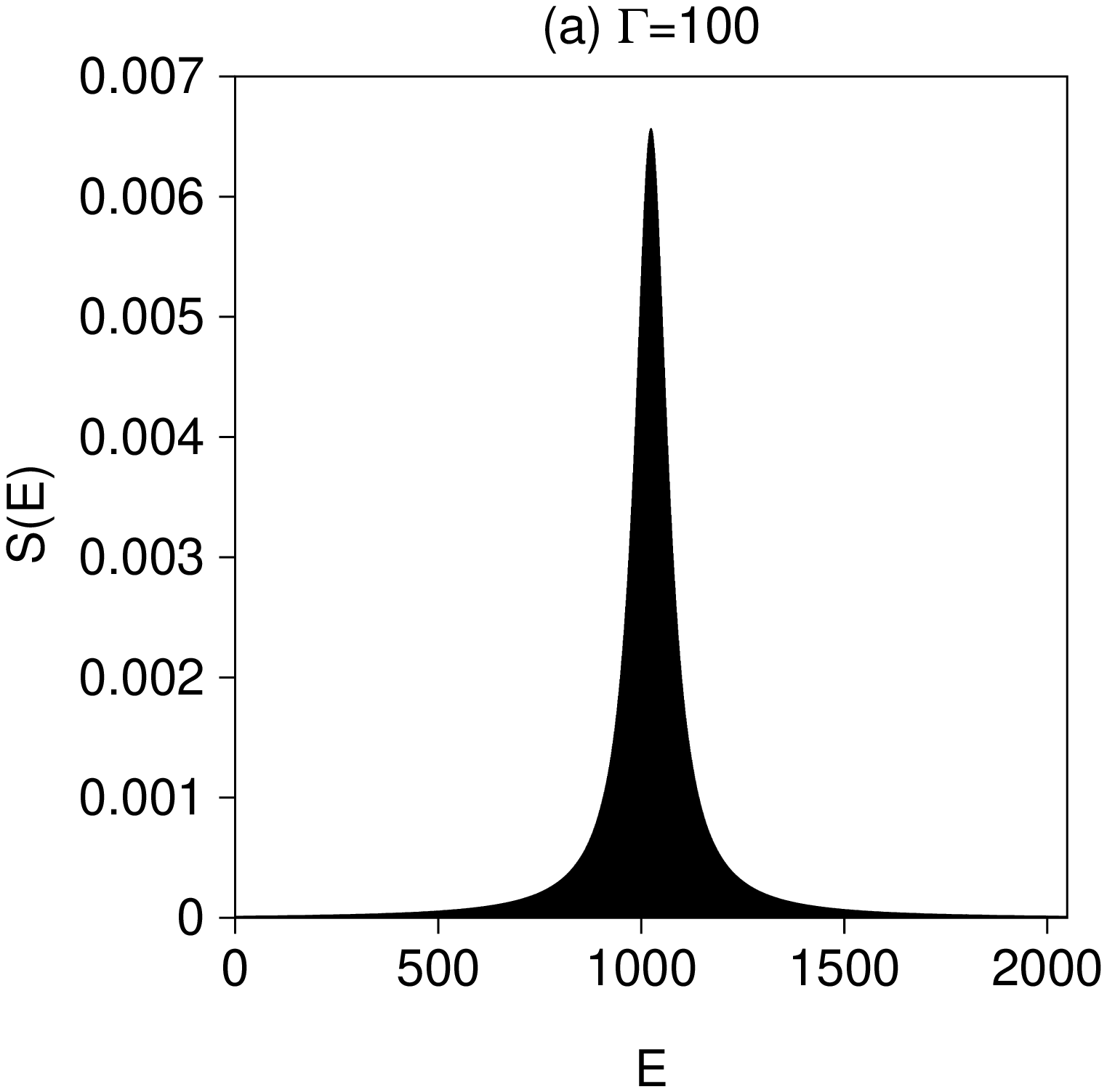,width=7.5cm}
  \end{minipage}
  \begin{minipage}{7.5cm}
       \psfig{file=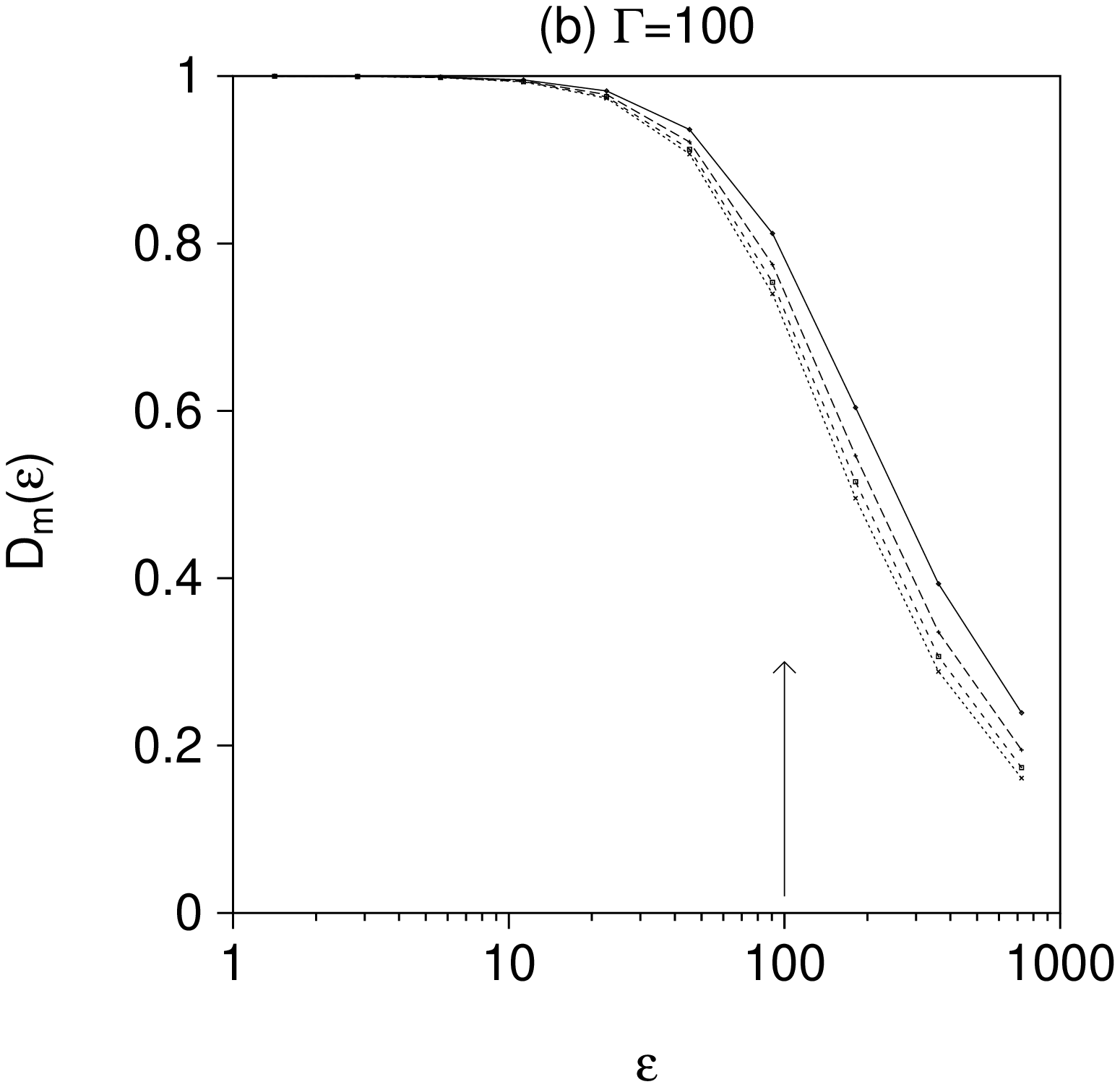,width=7.5cm}
  \end{minipage}
\end{center}
\caption{
(a)The Lorentzian strength function, and 
(b) the corresponding local scaling dimension $D_m(\epsilon)$.
The curves correspond to $D_m(\epsilon)$ for $m=2$ to 5 from upper to
lower.
The arrow indicates the value of the width $\Gamma$ of the Lorentzian.
}
\label{figloren}
\end{figure}

As the first example, we consider the Lorentzian distribution
\begin{equation}
S(E)= {1\over 2\pi}\frac{\Gamma}{E^2 + \Gamma^2/4}.
\end{equation}
This represents a typical peak profile associated with many kinds of
damping processes. 
The Lorentzian distribution shows a smooth variation of the strength
centered around the peak.
Discreteness of the energy levels is neglected.
Note that the Lorentzian distribution has one characteristic
energy scale which is the FWHM of the distribution, $\Gamma$.
The local scaling dimension $D_m(\epsilon)$ 
associated with the Lorentzian is shown in Fig.\ref{figloren}
as a function of the binning energy scale $\epsilon$. When
the bin width $\epsilon$ is smaller than
$\Gamma$, the local scaling dimension $D_m(\epsilon)$ takes the value
close to 1,  while $D_m(\epsilon) \sim 0$ for $\epsilon$
much larger than $\Gamma$. 
For $\epsilon$ around $\Gamma$, 
$D_m(\epsilon)$ decreases sharply with increasing $\epsilon$. 
This behavior arises because continuous and smooth aspect
of the Lorentzian distribution is dominantly measured
by the scaling analysis using small bin width $\epsilon \ll \Gamma$,
and the local scaling dimension goes to the
limit of uniform distribution $D_m(\epsilon)=1$.
With large bin width $\epsilon \gg \Gamma$, most of strength
is concentrated in a single bin, thus the local scaling dimension
reaches to the sharp peak limit 
$D_m(\epsilon)= 0$. With the bin width $\epsilon$ being the same order 
of the width $\Gamma$ of the Lorentzian, a transition between the
two limits occurs. Thus 
the profile of $D_m(\epsilon)$ relates to the characteristic
energy scale $\Gamma$ of the Lorentzian. In fact the value of
$\epsilon$ where $D_m(\epsilon)$ decreases most steeply 
corresponds well to $\Gamma$ as seen in Fig. \ref{figloren}.

\subsubsection{GOE strength fluctuation}
\label{sec:GOE}

If we distinguish the strengths associated with individual energy levels
of highly excited nucleus, the strengths fluctuate from
level to level on top of a smooth profile of strength distribution.
The level-by-level fluctuation originates 
from the chaotic nature of excited nuclei and is known to 
follow generic statistical rules described
by the random matrix theory of the Gaussian orthogonal 
ensemble (GOE) \cite{brody,dyson,mehta,bohigas}.
Let us characterize the GOE strength fluctuation by means of the scaling
analysis.

When the strength fluctuation is described by the GOE, 
the strengths $S_i$ for individual states fluctuate independently
and their statistics follows the Porter-Thomas 
distribution \cite{porter,brody} at large $N_{\rm tot}$ 
limit ($N_{\rm tot}$ being the dimension of the random matrix). 
The distribution of
energy levels $E_i$ follows a semi-circle level
density in average, and the position of the individual energy levels 
fluctuates locally around the average.
We treat here
only the strength fluctuation by neglecting effects 
both of the average level distribution and of local energy 
level fluctuations. Namely we
assume the uniform distribution of the energy levels. 
As discussed in Appendix\ \ref{appb}, influence of
the local energy level fluctuation on the local scaling dimension
is small due to the spectral rigidity of the GOE energy levels.
In this treatment, the
energy levels are chosen to be equally spaced: $E_i = id~
(i=1,2,...,N_{\rm tot})$, where $d$ is the level spacing.
The local scaling dimension  is then given by
\begin{equation} 
D_m(l)={1\over m-1}\sum_{i=1}^{m-1}{l\over l+2i},
\label{dimgoederriv}
\end{equation}
where $l\equiv\epsilon/d$ (the bin width measured in the unit
of the level spacing $d$) is used as the scaling parameter.
We give its derivation in Appendix\ \ref{appb}.

\begin{figure}
\begin{center}
  \begin{minipage}{7.5cm}
       \psfig{file=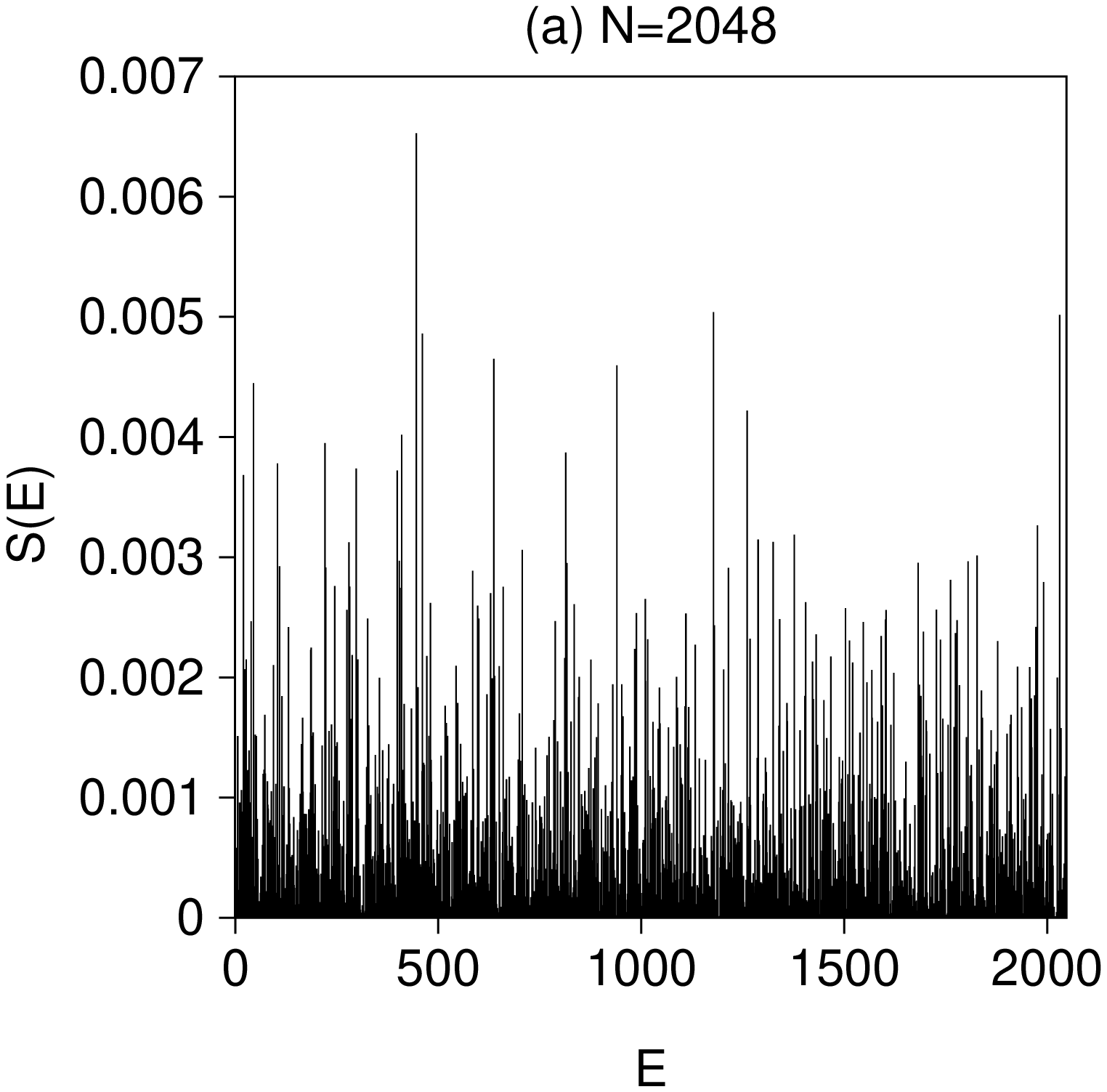,width=7.5cm}
  \end{minipage}
  \begin{minipage}{7.5cm}
       \psfig{file=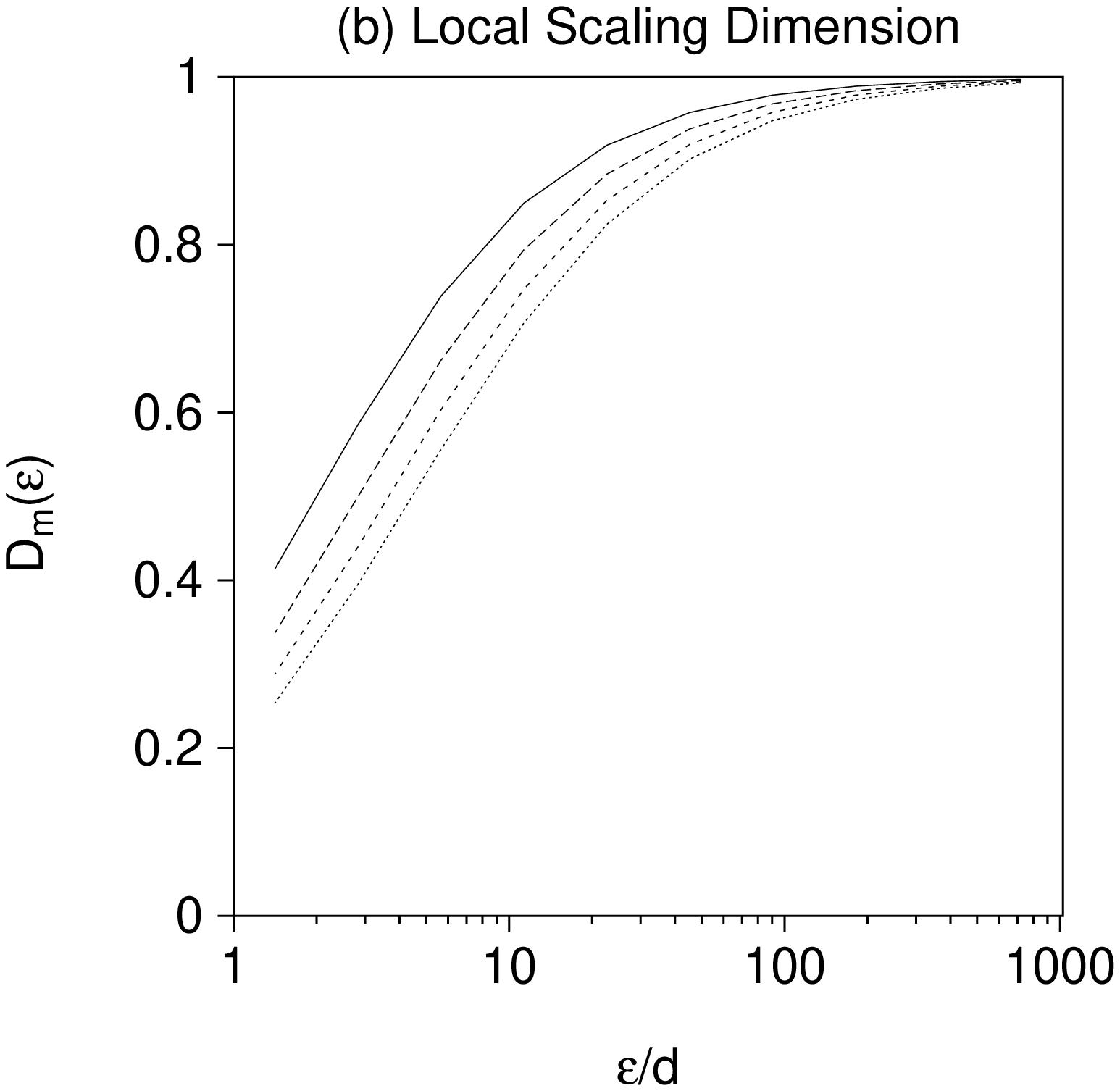,width=7.5cm}
  \end{minipage}
\end{center}
\caption{
(a)An example of the GOE strength function with matrix dimension of 2048.
(b)The local scaling dimension $D_m(\epsilon)$ for the GOE strength
 function obtained after the ensemble average. 
}
\label{figgoe}
\end{figure}

An example of the GOE strength function is shown in
Fig.\ \ref{figgoe}(a). The strengths of individual
states fluctuate strongly, i.e., there exist  levels which have 
as large strength as factor tens of the average value of strength.
This is because the Porter-Thomas distribution is a distribution
with large skewness.
Because of the strength fluctuation, the binned strengths
also fluctuate when the bin width $\epsilon$ is not very large
compared with the level spacing $d$. 
The fluctuation of the binned strength $\{p_n\}$ decreases as $\epsilon$
increases since the binning 
operates as an averaging over levels included in bins. 

The local scaling dimension $D_m(l)$, Eq.\ (\ref{dimgoederriv}),
is plotted in Fig.\ \ref{figgoe}(b), which shows 
how fluctuation of the binned
strength distribution changes as the energy scale (bin width) 
$\epsilon$ changes. 
When we use the bins whose width is comparable with the level
spacing ($l \lesssim 10$), the binned distribution shows many
spiky peaks caused by  large $S_i$ components. This leads
to small values of $D_m(\epsilon)$. The fluctuation of the
binned distribution is still large for the bin width of
order of ten level spacings ($l \sim 10^1$), where  
$D_m(\epsilon)$ deviate significantly from the uniform limit
$D_m(\epsilon)=1$. It is washed out mostly only with large bin width
of order of $100d$, where $D_m(\epsilon) \approx 1$ is realized.
It is also noted that the local scaling dimension $D_m(\epsilon)$
for the higher moment (larger $m$) takes significantly smaller
values compared with the second moment ($D_2(\epsilon)$) especially
for the small bin widths $l \lesssim 10^1$. Since taking the higher moment
emphasizes large components of the binned distribution $\{p_n\}$,
the local scaling dimension $D_m(\epsilon)$ for higher $m$ reflects
scaling property of those components. Small value of $D_m(\epsilon)$
for large $m$ indicates that the large components are distributed
sparsely along the $E$ axis, reflecting the spiky distribution of the
GOE strength.

We evaluated also the local scaling dimension $D_m(\epsilon)$
based on the finite difference definition, 
both by using the analytic evaluation Eq.\ (\ref{dimgoedif})
and by using numerical evaluation 
adopting 60 realizations of GOE.
They give essentially 
the same curves as the those plotted in Fig.\ \ref{figgoe}(b)
based on Eq.\ (\ref{dimgoederriv}), and
the difference is almost invisible in this plotting scale. 

When we include the energy level fluctuation, the local scaling
dimensions is affected for small bin width $\epsilon$ comparable to
the level spacing. However, the effect is not very large and becomes
negligible for $\epsilon > 10d$, as discussed in Appendix\ \ref{appb}.

\subsubsection{Poisson fluctuation}

As the third example, we consider the Poisson fluctuation.
If we consider experimental spectra where events are
counted in channels which corresponds to excitation energy bins, 
there always exists fluctuation in the spectra that originates
from the counting statistics. Even if the strength distribution
is completely uniform, counts in bins fluctuate statistically
obeying the Poisson distribution. 
Relating the
normalized `strength' $p_n$ to the counts $\{ r_n \}$ in
bins by $p_n= r_n/N$ ($N$ being the total counts), 
the partition function $\chi_m$ is expressed in 
terms of the $m$-th moment of the Poisson distribution.
Using the average number of counts $l=\langle r_n
\rangle$ as the scale of the bin width, the local scaling
dimension $D_m$ is given by
\begin{eqnarray}
D_m(l)&= &{1\over(m-1)}\frac{d\log(\langle r_n^m\rangle/l)}{d\log l}
\nonumber \\
      &= &{l \over l+1},~ {l^2+{3\over2}l \over l^2 + 3l+1},~
{l^3+4l^2+{7\over3}l \over l^3+6l^2+ 7l+1},~
{l^4+{15\over2}l^3+{25\over2}l^2+{15\over4}l \over l^4+10l^3+25l^2+
      15l+1} \ \ (m=2,3,4,5).
\label{dmpoisson}
\end{eqnarray}
The local scaling dimension takes the value significantly smaller than one
for small $l$ ($l \lesssim 10$). It monotonically increases with
increasing $l$ and approaches to one.  
The behavior is similar to that of
the GOE strength fluctuation, Fig.\ref{figgoe}(b), 
but the values of $D_m$ for the Poisson fluctuation is closer to 
$D_m=1$ than the GOE fluctuation since the fluctuation associated with
the Poisson distribution is smaller than the 
GOE fluctuation. 

Expression\ (\ref{dmpoisson}) indicates that effects of the counting
statistics diminish for $l$ larger than several tens for which
the uniform limit $D_m \approx 1$ is almost achieved.
We neglect the counting statistics in the other part of this paper.

\subsection{Relation to Autocorrelation Analysis}

One often uses the autocorrelation function to characterize
fluctuations of the strength function.
The autocorrelation function for the strength
function,  defined by 
\begin{eqnarray}
C_2(\epsilon) &\equiv &\int S(E)S(E+\epsilon)dE,\\
\label{def1c2}
&=&\sum_{i,j}S_iS_j\delta(\epsilon-E_j+E_i),
\label{def2c2}
\end{eqnarray}
quantifies the fluctuation correlation as a function of the
displacement energy $\epsilon$. Dependence of $C_2(\epsilon)$
on the displacement energy $\epsilon$ reveals the characteristic
energy scale involved in the strength fluctuations. Since the
local scaling dimension has a similar property as discussed above,
one may expect relation between the autocorrelation function and the
local scaling dimension.
In fact, we can prove that $C_2(\epsilon)$ is related to 
the partition function $\chi_2(\epsilon)$ and the local scaling dimension
$D_2(\epsilon)$ for the second moment. 

To this end, we introduce an integral function of $C_2(\epsilon)$
by 
\begin{equation}
B_2(\epsilon)\equiv \int_{-\epsilon}^\epsilon C_2(\epsilon')d\epsilon' \\
   =\sum_{ij}S_iS_j\theta(\epsilon-|E_i-E_j|).  
\label{b2theta}
\end{equation}
Comparing Eq.\ (\ref{b2theta}) with the closed form expression Eq.\
(\ref{inichi2theta}) for the partition function, 
we find an equation
\begin{equation}
(1+\epsilon{d\over d\epsilon})\chi_2(\epsilon)
=B_2(\epsilon).
\label{derivchi2b2}
\end{equation}
Thus, the partition function is expressed as 
\begin{equation}
\chi_2(\epsilon)={1\over\epsilon}\int_0^\epsilon
B_2(\epsilon')d\epsilon',
\label{chi2b2}
\end{equation}
in terms of the integral of $B_2(\epsilon)$.

Using this expression for Eq.\ (\ref{scaledim}), the
local scaling dimension $D_2(\epsilon)$ is 
written as
\begin{equation}
D_2(\epsilon)=\frac{B_2(\epsilon)-\chi_2(\epsilon)}{
\chi_2(\epsilon)},
\label{dmbm}
\end{equation}
which is expressed in terms of
the integral  $B_2(\epsilon)$ and the double integral
$\chi_2(\epsilon)$ of the
autocorrelation function  $C_2(\epsilon)$. 

The close connection between the autocorrelation function 
$C_2(\epsilon)$ and the local scaling dimension $D_2(\epsilon)$ 
can be seen easily for the case of the Lorentzian distribution. 
For the Lorentzian distribution, the autocorrelation function 
$C_2(\epsilon)={1\over\pi}{\Gamma\over\Gamma^2 + \epsilon^2}$ decreases
monotonically with increasing $ \epsilon$ and its value changes
most steeply around $\epsilon \sim \Gamma$, where
$\Gamma$ is the FWHM of the Lorentzian strength distribution.
On the other hand, 
the local scaling dimension $D_2(\epsilon)$ is given with use of
Eq.\ (\ref{dmbm}) by
\begin{equation}
D_2(\epsilon) = \frac{{\Gamma\over 2\epsilon}\log((\epsilon/\Gamma)^2+1)}
{\tan^{-1}{\epsilon\over\Gamma} - 
{\Gamma\over 2\epsilon}\log((\epsilon/\Gamma)^2+1)}.
\end{equation} 
This function also decreases monotonically from 
$D_2=1$ at $\epsilon=0$ with increasing $\epsilon$.
The point  of its steepest slope, 
$\epsilon=0.80\Gamma$,
(or $\epsilon=1.8\Gamma$ when plotted as a function of $\log\epsilon$)
corresponds well to the FWHM of the Lorentzian.
This correspondence holds also for the local scaling dimensions
$D_m$ of higher moments as seen in the previous subsection.

There are two figures of merit for the use of the scaling analysis
in comparison with the autocorrelation analysis. Firstly,
the local scaling dimension $D_m(\epsilon)$ has
smooth $\epsilon$ dependence while the autocorrelation
function $C_2(\epsilon)$ is a sum of the delta functions (if we do not
include any smoothing procedure). This difference can be seen in Eq.
(\ref{dmbm}) where the $D_2(\epsilon)$ is expressed in terms of the
integral functions of $C_2(\epsilon)$. Secondly,
the local scaling dimensions $D_m(\epsilon)$ for $m>2$ carry
information on the higher moments of strength fluctuation.

\section{Analysis of Doorway Damping Model}
\label{sec:numerical}
In the following we shall discuss fluctuations of the strength
function which is associated with damping phenomena of a
specific state, e.g., a collective vibrational state, embedded
in the background states in the highly excited nuclei. 
Keeping in mind the damping mechanism of the giant resonances,
we herewith adopt a model system in which the damping 
of a collective state takes place through coupling to
doorway states which consist of only a part of the background states.
The strength function  of the collective state
exhibits not only spreading of the strength common for all
damping phenomena but also fine structures reflecting the
presence of the doorway states. We shall discuss in detail how
the scaling analysis describes characteristics of
this kind of strength fluctuations.

\subsection{Model}

The Hamiltonian of the model is given by
\begin{equation}
H=\omega_c|c\rangle\langle c|+V_{\rm doorway}+H_{\rm bg},
\label{hamiltonian}
\end{equation}
where  $|c\rangle$ denotes the collective state, 
whose energy is $\omega_c$.
The third term represents the Hamiltonian for the background states,
which include the doorway states as well as the other background states.
The second term $V_{\rm doorway}$  represents the coupling of the 
collective state $|c\rangle$ to the doorway states.

Here we keep in mind the picture 
that shell model many-particle many-hole 
excitations form basis states $\{|\mu\rangle\}$ of the background states
and
the interaction among the basis states causes configuration mixing.
As a simplified treatment, we assume that $H_{\rm bg}$ has diagonal
term representing the unperturbed energies of the basis states, and for the
interaction among the basis states we employ a GOE Hamiltonian. Namely,
\begin{equation}
H_{\rm bg}=\sum_\mu\omega_\mu|\mu\rangle\langle \mu|+H_{\rm GOE},
\label{bghamil}
\end{equation}
\begin{equation}
H_{\rm GOE}=\sum_{\mu} v^{\rm d}_{\mu\mu}|\mu\rangle\langle\mu|+
\sum_{\mu > \nu}v^{\rm nd}_{\mu\nu}(|\mu\rangle\langle \nu|+h.c.).
\label{goehamil}
\end{equation}
The basis energy $\omega_\mu$ is given by an equidistant model,
\begin{equation}
\omega_\mu=(-{N_{\rm bg}\over 2}+\mu)d~~~(\mu=1,\cdots,N_{\rm bg}),
\label{bgenergy}
\end{equation}
where $d$ is the level spacing of the background states and
$N_{\rm bg}$ represents the total number of backgrounds.
Matrix elements $v^{\rm d}_{\mu\mu}$ and $v^{\rm nd}_{\mu\nu}$
($\mu\neq \nu$) for the GOE Hamiltonian
are independent Gaussian random variables with the zero mean and 
the variance satisfying 
$\langle (v^{\rm d}_{\mu\mu})^2\rangle =
2 \langle (v^{\rm nd}_{\mu\nu})^2\rangle$.
Because of the interaction, the background states have
their own spreading width, denoted $\gamma$ hereafter, which
can be estimated as,
\begin{equation}
\gamma=2\pi{\langle (v^{\rm nd}_{\mu\nu})^2\rangle \over d}.
\label{bggamma}
\end{equation}
The background Hamiltonian\ (\ref{bghamil}) is 
equivalent to the model adopted in ref.\ \cite{guhr}
and the same as the Wigner ensemble random matrix model\
\cite{wigner}.

The second term in Eq.\ (\ref{hamiltonian}) representing the coupling
between
the collective state and doorway states is given by

\begin{equation}
V_{\rm doorway} = \mathop{{\sum}'}_{\mu={\rm doorway}} V_\mu^{\rm door}
(|c\rangle\langle \mu|+h.c.).
\label{doorwaycouple}
\end{equation}
Here the interaction is present only for a part of the
background $\mu$ states which are supposed to be the doorway states of the
damping, and the summation in Eq.(\ref{doorwaycouple}) runs only over
the doorway states. The doorway $\mu$ states 
are selected in every $L$ $\mu$-states (i.e., $\mu$ states with 
$\mu=L,2L,3L,...$).
Accordingly, the level spacing of doorway states is
$D=Ld$. The coupling matrix elements  $V^{\rm door}_\mu$ 
are random variables taken from a Gaussian distribution with zero mean.

By diagonalizing the Hamiltonian with use of the basis consisting of
all the $\mu$ states and the collective state $|c\rangle$, we obtain for
each
realization of random variables the
energy eigenstates $|i\rangle$, their energy $E_i$ and
associated strength $S_i=|\langle c|i \rangle|^2$ of the collective state.
The strength function of the collective state
\begin{equation}
S(E)=\sum_i S_i\delta(E-E_i)
\label{strfun}
\end{equation}
is thus obtained. Since the collective-doorway couplings are uniform 
with respect to the energy of the background states, the
strength function exhibits as a global profile the Lorentzian
distribution. In fact, when we average the strength
distribution over many realizations of the Hamiltonian and make
smoothing with respect to the energy $E$, the averaged strength function
 $\tilde{S}(E)$ becomes very close to the Lorentzian whose FWHM is
given by the golden rule estimate
\begin{equation}
\Gamma=2\pi{\langle (V^{\rm door}_\mu)^2\rangle \over D}. 
\label{gamma}
\end{equation}
In this paper we do not discuss the global spreading
profile since it is trivial in this model and it often
has simple profile such as Lorentzian also in the 
experimental data. Instead
we deal only with fine structures and fluctuations of
the strength function. It is then useful to 
remove the global smooth profile by normalizing
the strength $S_i$ as
\begin{equation}
\bar{S}_i ={\cal N}{S_i\tilde{\rho}(E_i) \over
\tilde{S}(E_i)}.
\label{unfold}
\end{equation}
Here, 
$\tilde{S}(E)$ and $\tilde{\rho}(E)$ are obtained by averaging 
the calculated strength function, Eq.(\ref{strfun}),
and the level density $\rho(E)=\sum_i\delta(E-E_i)$ over realizations
 and by smoothing with respect to $E$ with use of 
the Gaussian weighting and the Laguerre polynomials as adopted in
the Strutinsky method\ \cite{ring-schuck}.
$\cal{N}$ is an overall normalization factor 
to guarantee $\sum_i\bar{S}_i=1$.
As we discussed in connection
with the GOE strength fluctuation (Sec.\ \ref{sec:GOE} and
Appendix\ \ref{appb}), effect of the energy
level fluctuation on the fluctuation of the strength function
is small except for very small energy scale less than about
ten times of the level spacing $d$. 
Furthermore the level fluctuations within a small energy
interval is described by the GOE model as far as $\gamma$ is not small.
Thus we neglect the fluctuation of the energy levels in the following
analysis. In other words, we use the level order $i$ as the
scale of the excitation energy of the levels 
instead of the energy $E_i$ itself.
This is equivalent to assume that the energy levels are equidistantly
distributed as $E_i = id$.
In the following, we treat unfolded strength function
$\bar{S}(E)=\sum_i\bar{S}_i\delta(E-E_i)$ thus obtained.

The number of background states is fixed to $N_{\rm bg}=8191$ so
that the size of the Hamiltonian matrix has the dimension of 8192.
The unperturbed energy of the collective state is placed at the
energy center of the spectrum by putting $\omega_c=0$, 
and we set the FWHM of the global strength distribution  
$\Gamma=2000d$ so that the most of the strengths are located
dominantly in the central region of the total spectra. For the
interaction matrix elements in $H_{\rm bg}$, or equivalently, 
the spreading width $\gamma$ of the background states, we use
several different values $\gamma=64d$, $128d$, $256d$, and $512d$
to see the dependence on $\gamma$. The spacing $D$ between
the doorway states is also varied as  $D=64d$, $128d$, and $256d$, 
corresponding to the number of doorways 128, 64, 32, respectively.

By generating random numbers  for the Gaussian variables in the
Hamiltonian, 
a realization of the Hamiltonian is obtained. The spectra
and the strength distribution are calculated for each realization. 
We perform
an average over many realizations of the Hamiltonian to obtain the
features which is independent of specific realizations. For this purpose
we generate repeatedly the random numbers in $V_{\rm doorway}$, but 
fix the coefficients in $H_{\rm bg}~ (H_{\rm
GOE})$ since the results do not depend very
much on realizations of $H_{\rm GOE}$ because of 
sufficiently large dimension $N_{\rm bg}$.
For the ensemble average, we perform
averaging over 60 realizations, which is enough to obtain 
almost exact results for the GOE fluctuations (Sec.\ \ref{sec:GOE}).

The partition function is calculated for the bin widths of
powers of 2 of the level spacing $d$, i.e., $\epsilon=ld$, 
$l = 1,2,4,8,16,...,4096$. 
Since the number of levels is finite, we need to take care of
the edges of the strength distribution.
The bins placed at either edge of the spectra 
have smaller bin width than those in the other  part of the spectra,
and this would affect the partition function when the bin width is not very
small compared to the whole spectrum width. 
To avoid the edge effect, we assumed a periodic continuation of the
spectra by imposing 
$\bar{S}_{i+N_{\rm tot}}=\bar{S}_i$, where $N_{\rm tot}$ is
the total number of levels $N_{\rm tot}=N_{\rm bg}+1$.  For the
calculation of the partition function, we performed averaging over
16 bin boundary positions by sliding the bin boundaries  by $
l/16$ (we average over $l$ boundary positions for
 $l<16$), which provides sufficient averaging.

\begin{figure}
  \begin{minipage}{8cm}
       \psfig{file=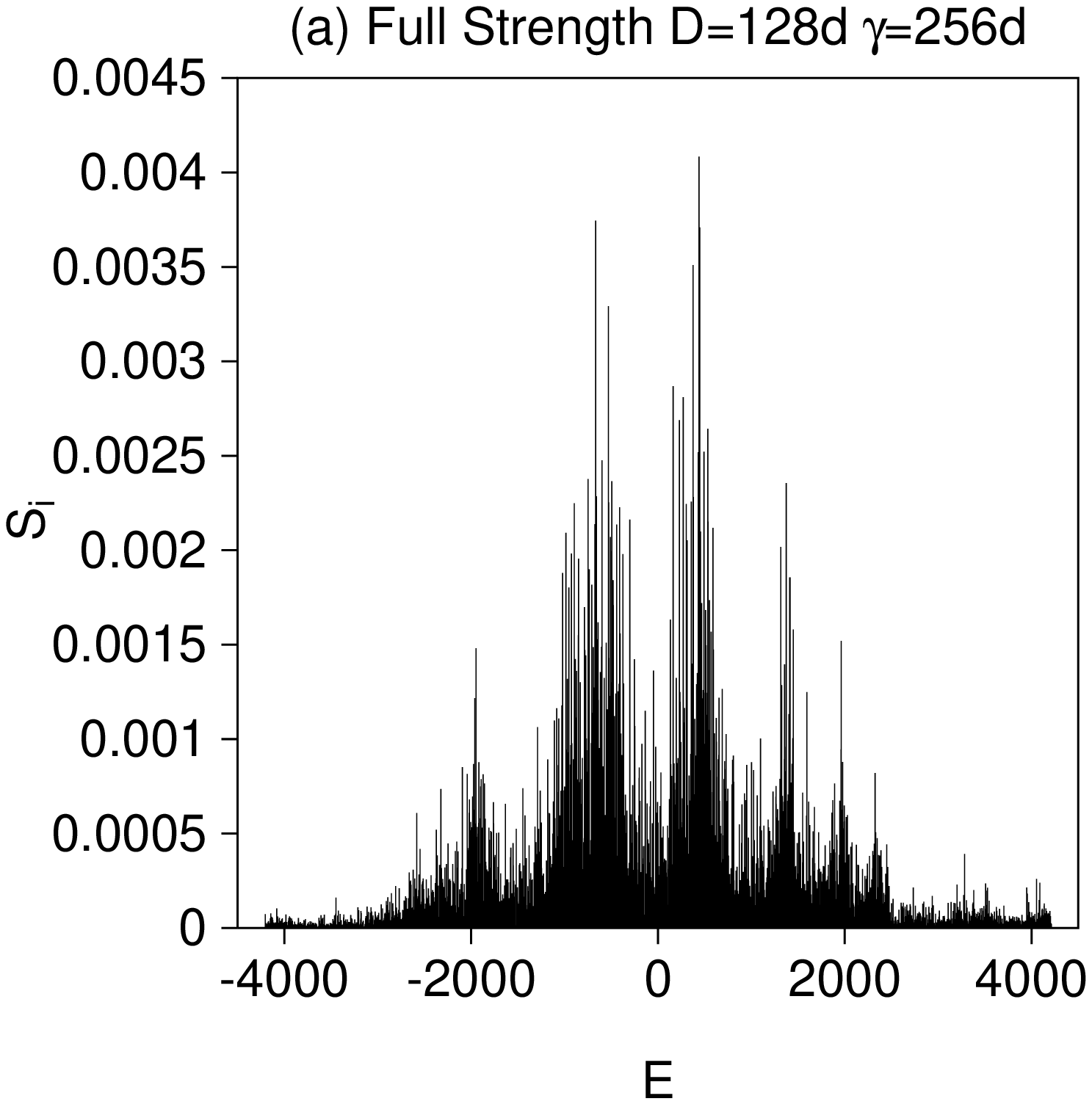,width=8cm}
  \end{minipage}
  \begin{minipage}{8cm}
       \psfig{file=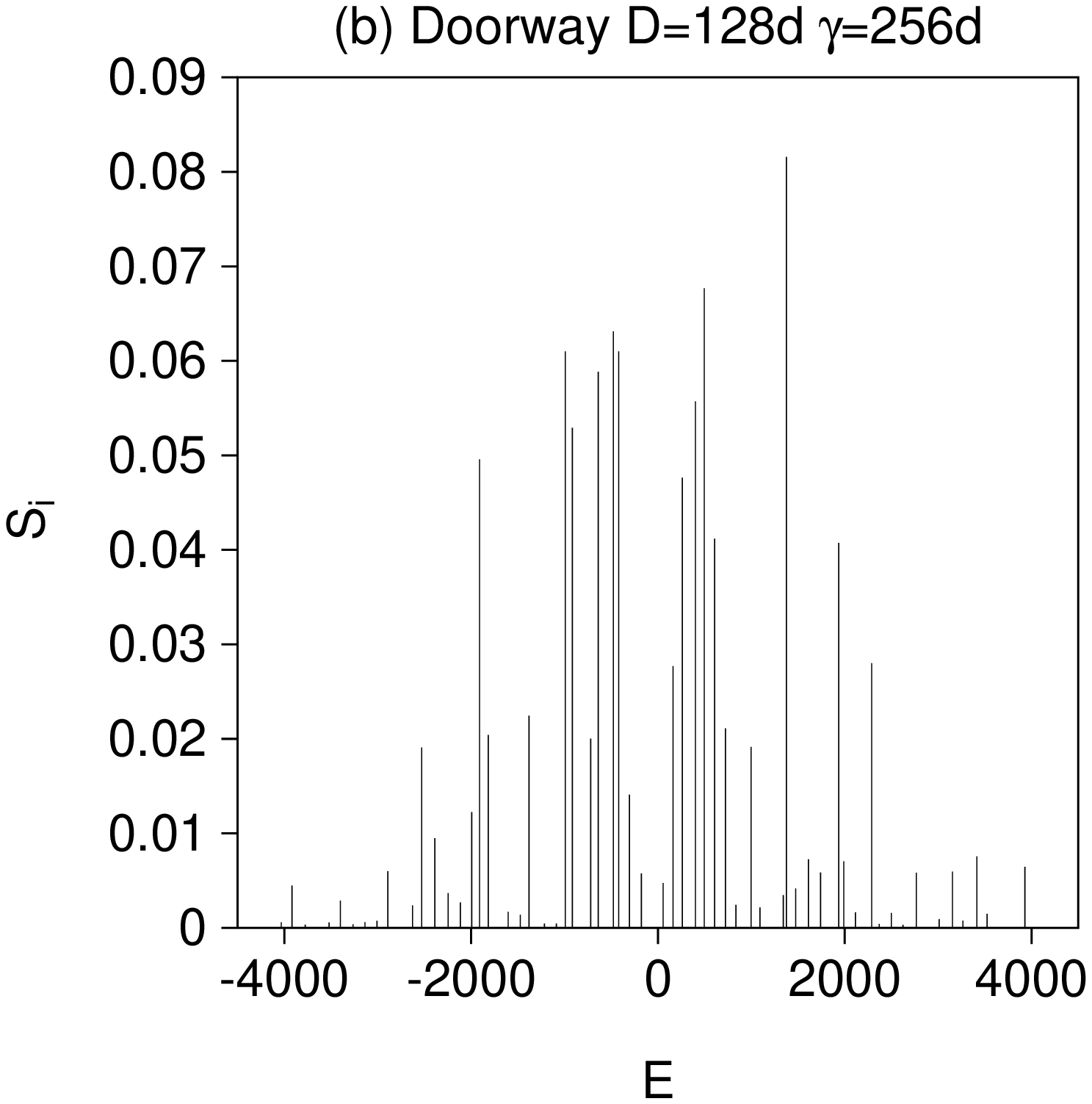,width=8cm}
  \end{minipage}
%\end{center}

%\begin{center}
  \begin{minipage}{8cm}
       \psfig{file=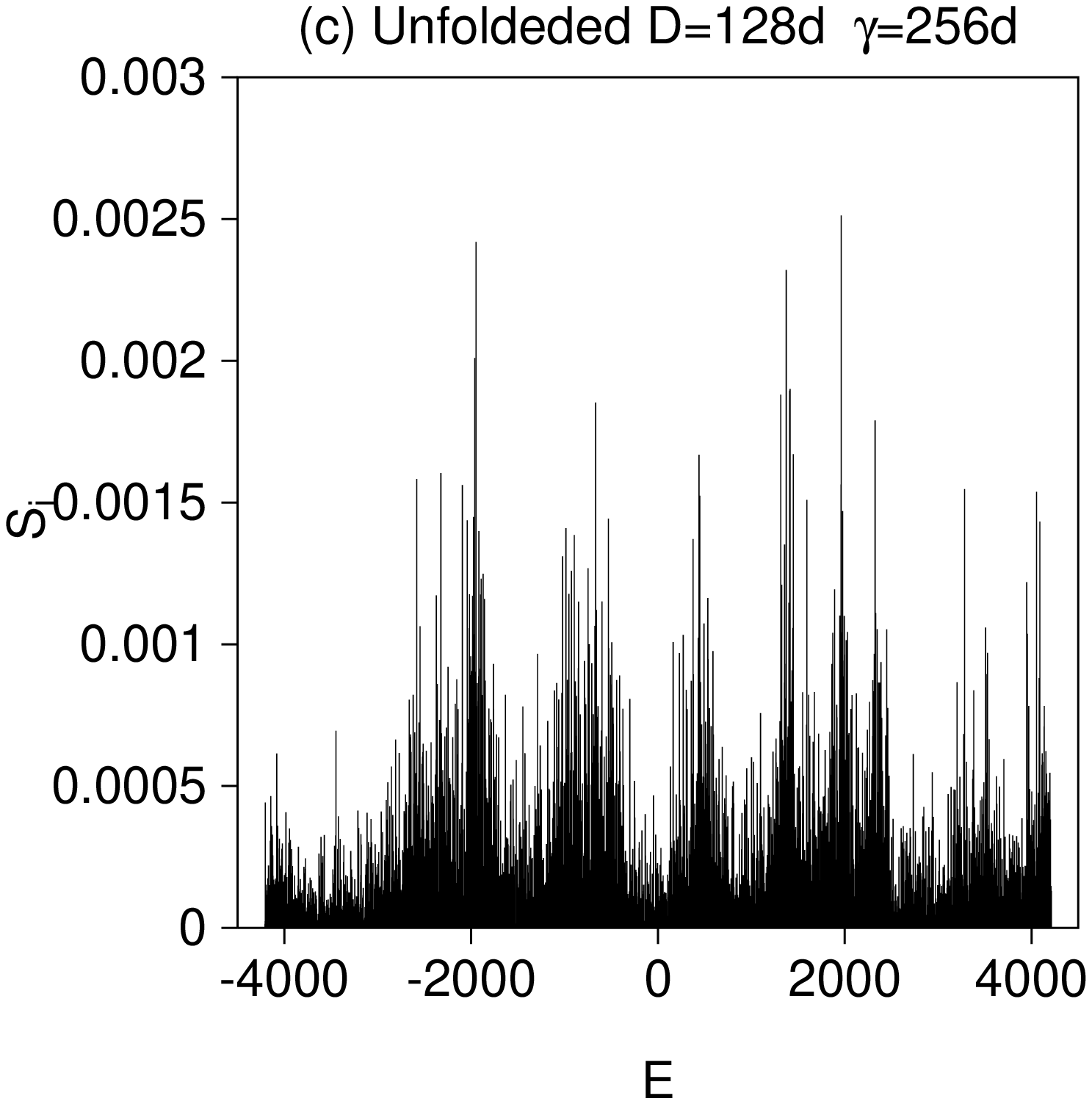,width=8cm}
  \end{minipage}
\caption{
(a)An example of the strength function obtained in the doorway damping
model with parameters $D=128d$ and $\gamma=256d$.
(b)The same as (a), but coupling between the doorway states and 
the other background states is neglected. 
(c)The unfolded strength function $\bar{S}(E)$ corresponding
to the original one plotted in (a).
}
\label{figvarstr}
\end{figure}

An example of the strength function is shown in Fig.\ \ref{figvarstr}(a).
Qualitatively the distribution looks a Lorentzian distribution
as a gross profile. However, the distribution has large fluctuation
around the Lorentzian shape.
In addition to apparent level-to-level fluctuations, one 
can recognize clustering of strengths. The latter  appears to reflect the
distribution of the collective-doorway couplings as seen from the
comparison
with Fig.\ \ref{figvarstr}(b), which shows the strength distribution
which is obtained by assuming pure doorway states, i.e., by neglecting
the interaction between the doorway states and the other background states.
Note however that it is very hard to identify 
individual doorway states in the strength distribution of Fig.\
\ref{figvarstr}(a)
since in this
example the spreading width of doorway states, $\gamma$, is larger than the
spacing between the doorways, $D$. The strengths associated with
individual doorways overlap with each other.
Fig.\ \ref{figvarstr}(c) shows the distribution of the normalized strength
(corresponding to Fig.\ \ref{figvarstr}(a)) for which we apply the scaling
analysis.

\begin{figure}
\begin{center}
  \begin{minipage}{8cm}
       \psfig{file=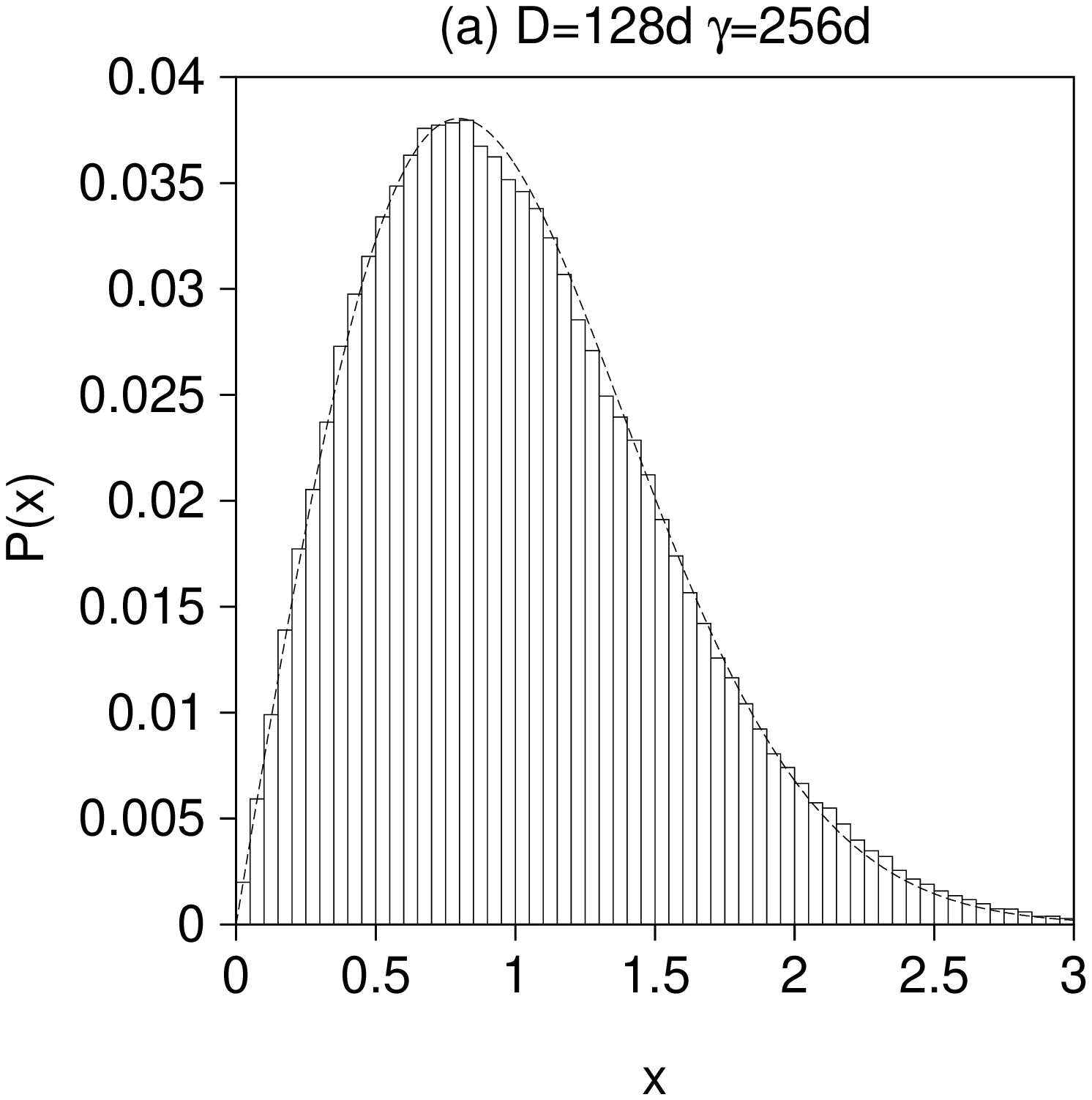,width=8cm}
  \end{minipage}
  \begin{minipage}{8cm}
       \psfig{file=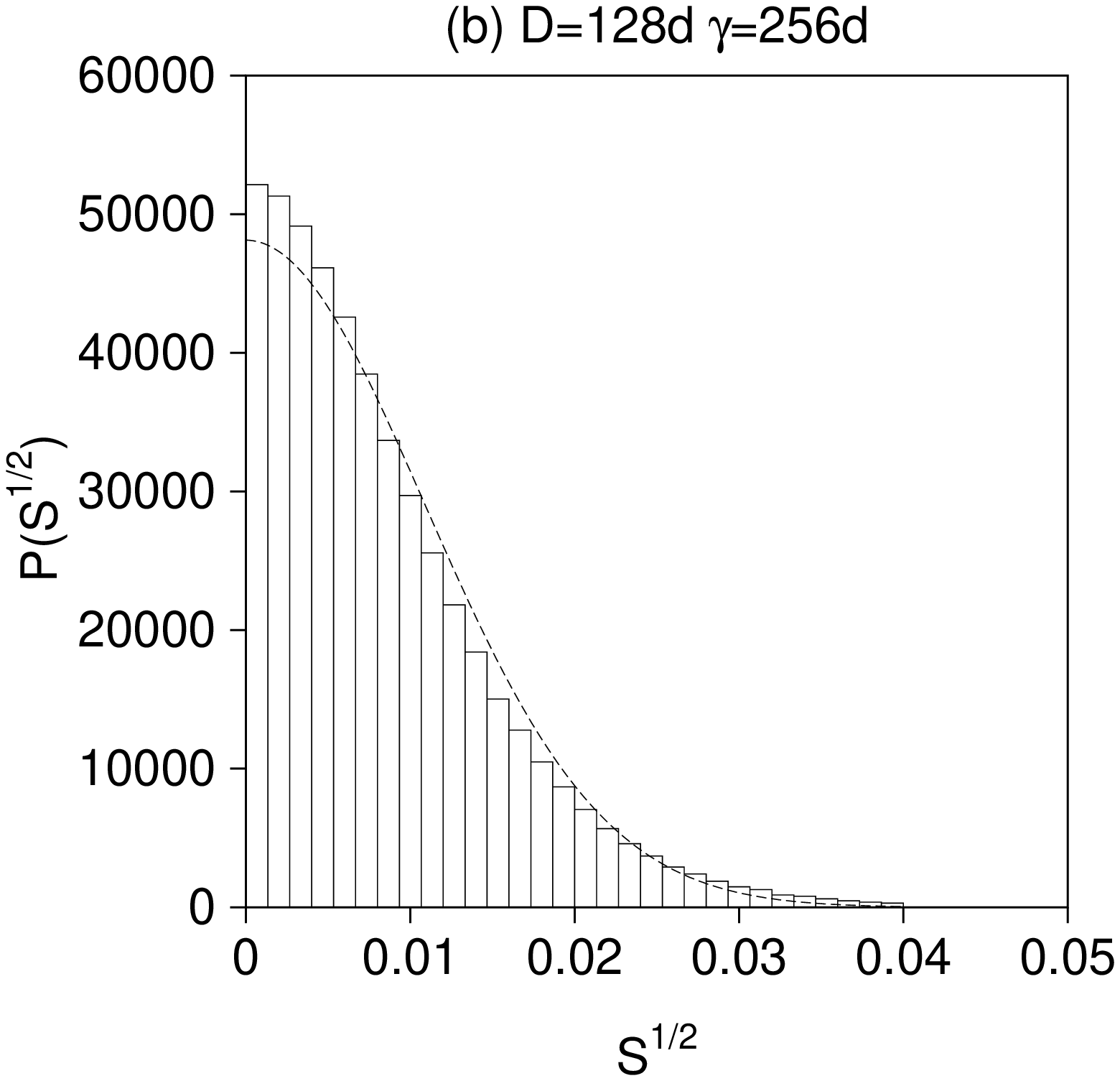,width=8cm}
  \end{minipage}
\end{center}
\caption{
(a)The nearest neighbor level spacing distribution for 
the doorway damping model with parameters $D=128d$ and $\gamma=256d$.
Level spacings were unfolded. The dashed curve represents
 the Wigner distribution.
(b)The statistical distribution of square root of 
unfolded strengths, $\bar{S}_i^{1/2}$,
associated with the individual levels for $D=128d$ and
$\gamma=256d$. The dashed curve represents the Porter-Thomas
distribution which becomes a Gaussian when plotted as a function
of $\bar{S}_i^{1/2}$.
 Both results are obtained with the ensemble average over 60
 realizations of the doorway
couplings.
}
\label{fignns}
\end{figure}

Fig.\ \ref{fignns}(a) shows the nearest neighbor level spacing distribution
(NND) for $D=128d$ and $\gamma=256d$. The NND follows
the Wigner distribution \cite{brody} perfectly, indicating that the
fluctuation of the energy level at a short energy interval
obeys the GOE random matrix limit.
By using the same Hamiltonian as our background Hamiltonian 
$H_{\rm bg}$, Guhr et al.\cite{guhr} showed that the GOE limit of the NND
is reached as the spreading width of the background states $\gamma$ 
becomes larger than several tens  of the level spacing $d$.   
Fig.\ \ref{fignns}(b) shows the statistical distribution 
of strength $\bar{S}_i$ of individual level for $D=128d$ and
$\gamma=256d$. Here the histogram of 
distribution of square root strength 
$\bar{S}_i^{1/2}$ is plotted
together with the Porter-Thomas distribution drawn
by the dashed curve. Note that the Porter-Thomas distribution 
$P(S)=\exp(-S/2)/\sqrt{2\pi S}$
 has a Gaussian shape 
in this plotting.
Deviation from the
Porter-Thomas distribution is clearly seen. 
The excess of large and small strengths
compared to the Porter-Thomas distribution indicates that 
the strength fluctuation is larger than the GOE limit. 
The deviation is caused by the clustering of the strength
distribution, which originates from the fine structures of the
strength functions associated with doorway states of the
damping of the collective state.

\subsection{Scaling Analysis}
\label{sec:results}

\subsubsection{Characteristics of $D_m(\epsilon)$}
Let us analyze the strength fluctuation by applying the scaling
analysis which we introduced in the previous section.

\begin{figure}
\begin{center}
  \begin{minipage}{8cm}
       \psfig{file=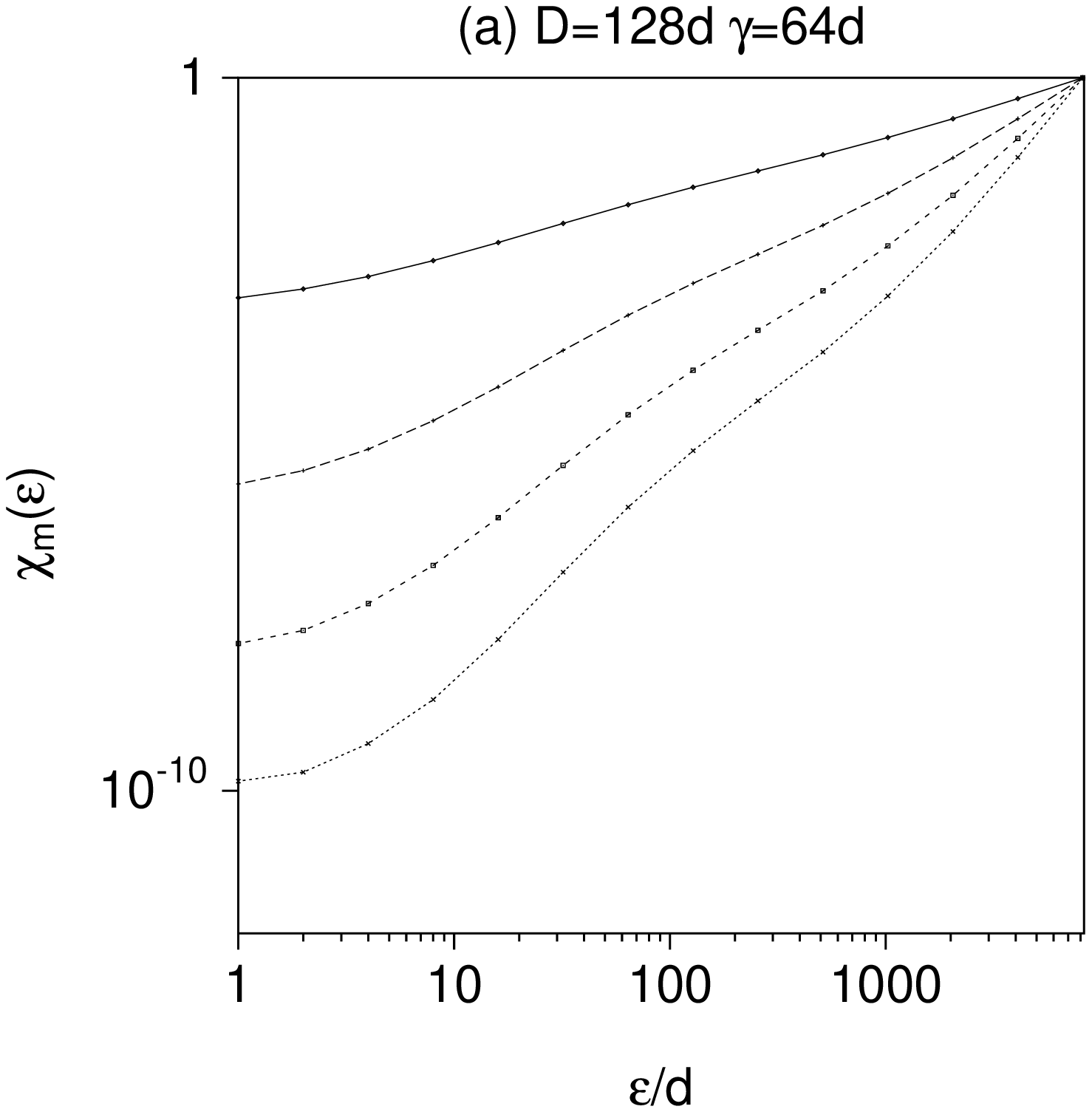,width=8cm}
  \end{minipage}
  \begin{minipage}{8cm}
       \psfig{file=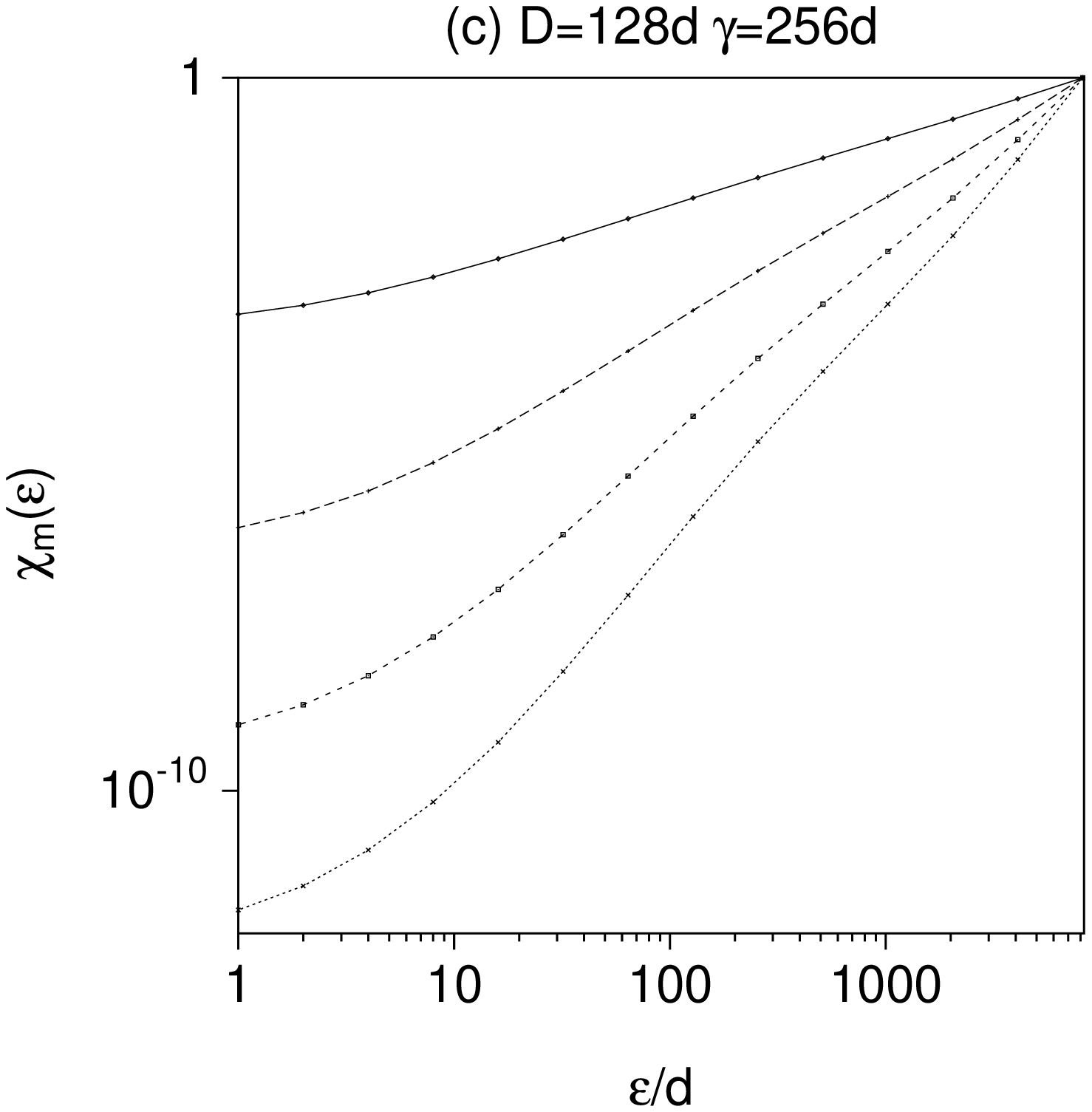,width=8cm}
  \end{minipage}
\end{center}
\begin{center}
  \begin{minipage}{8cm}
       \psfig{file=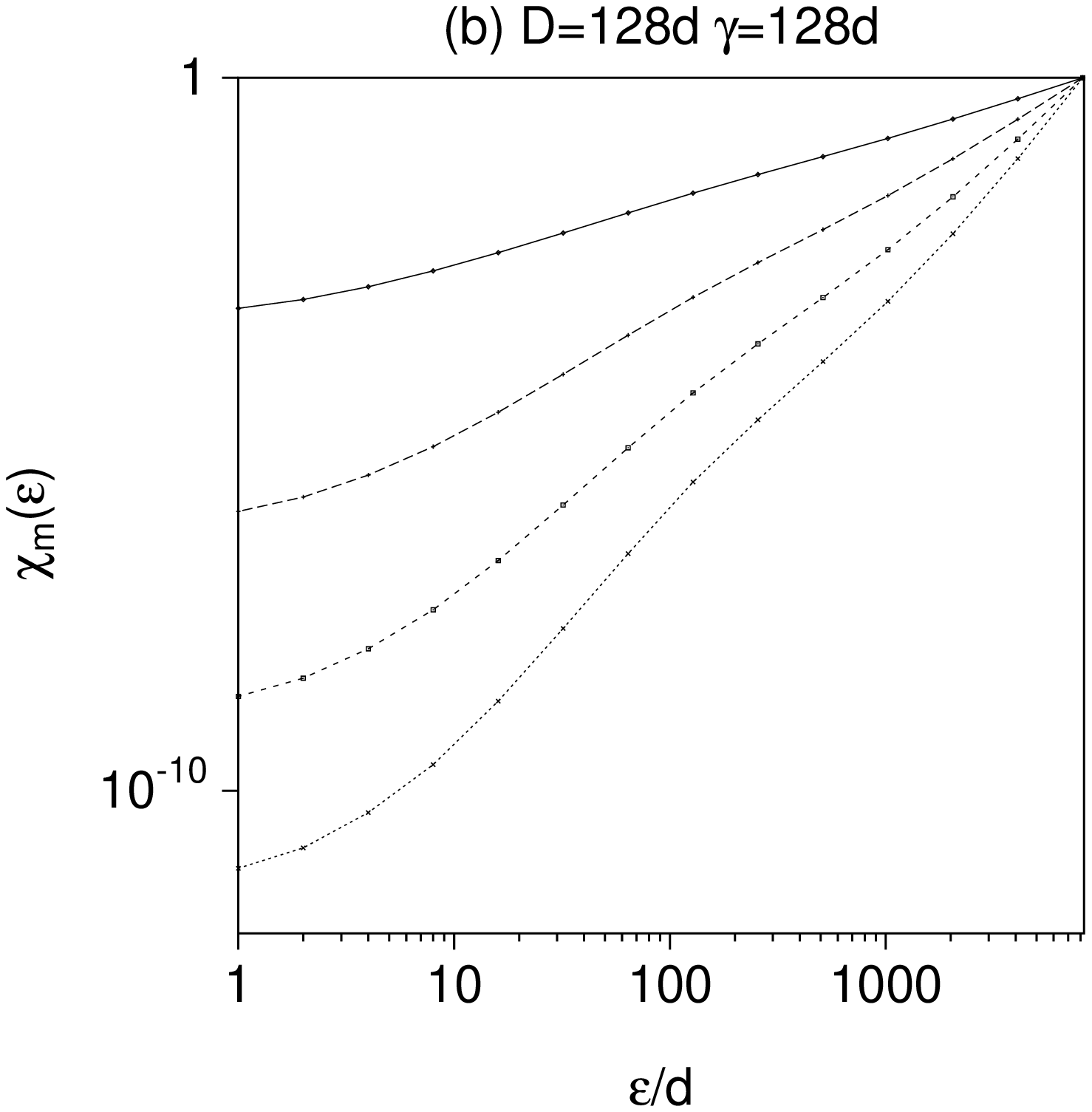,width=8cm}
  \end{minipage}
  \begin{minipage}{8cm}
       \psfig{file=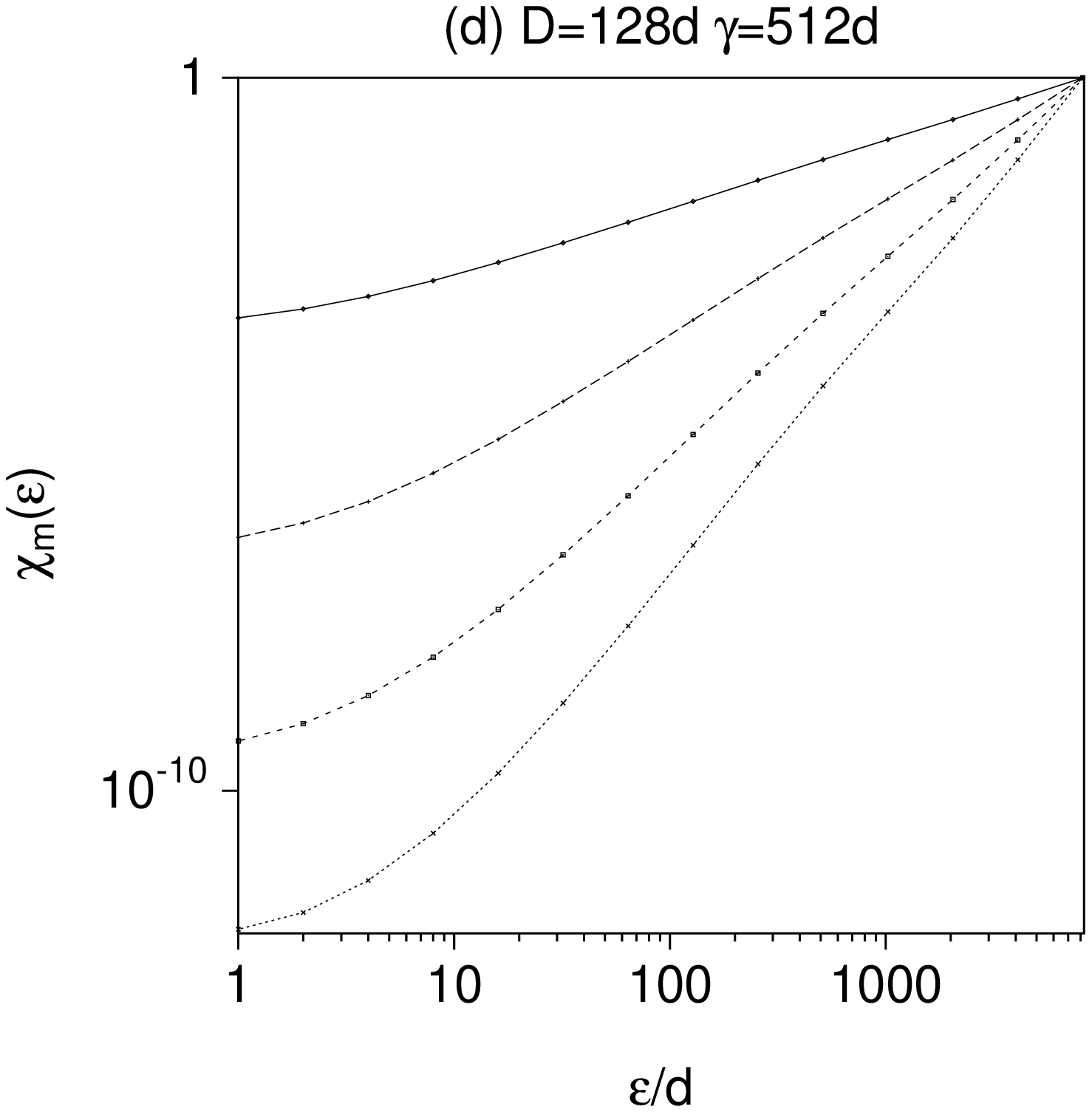,width=8cm}
  \end{minipage}
\end{center}
\caption{
The partition function $\chi_m(\epsilon)$ for the 
doorway damping 
model with four different values of
(a) $\gamma=64d$, (b) $128d$, (c) $256d$, and (d) $512d$ while
$D=128d$ is fixed. The result is obtained with the ensemble average
over 60 realizations of the doorway 
couplings.
}
\label{figpartition}
\end{figure}

\begin{figure}
\begin{center}
  \begin{minipage}{8cm}
       \psfig{file=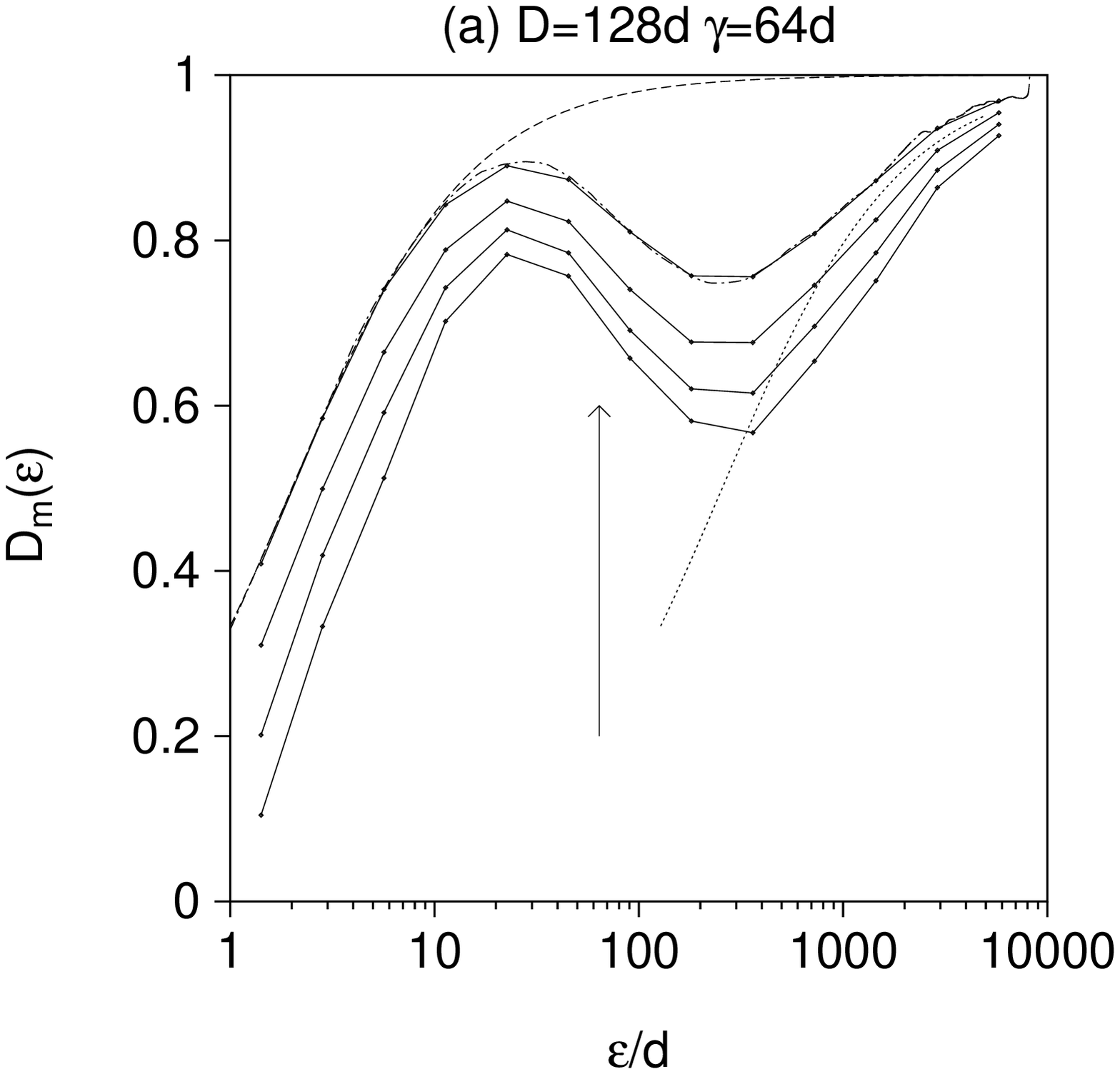,width=8cm}
  \end{minipage}
  \begin{minipage}{8cm}
       \psfig{file=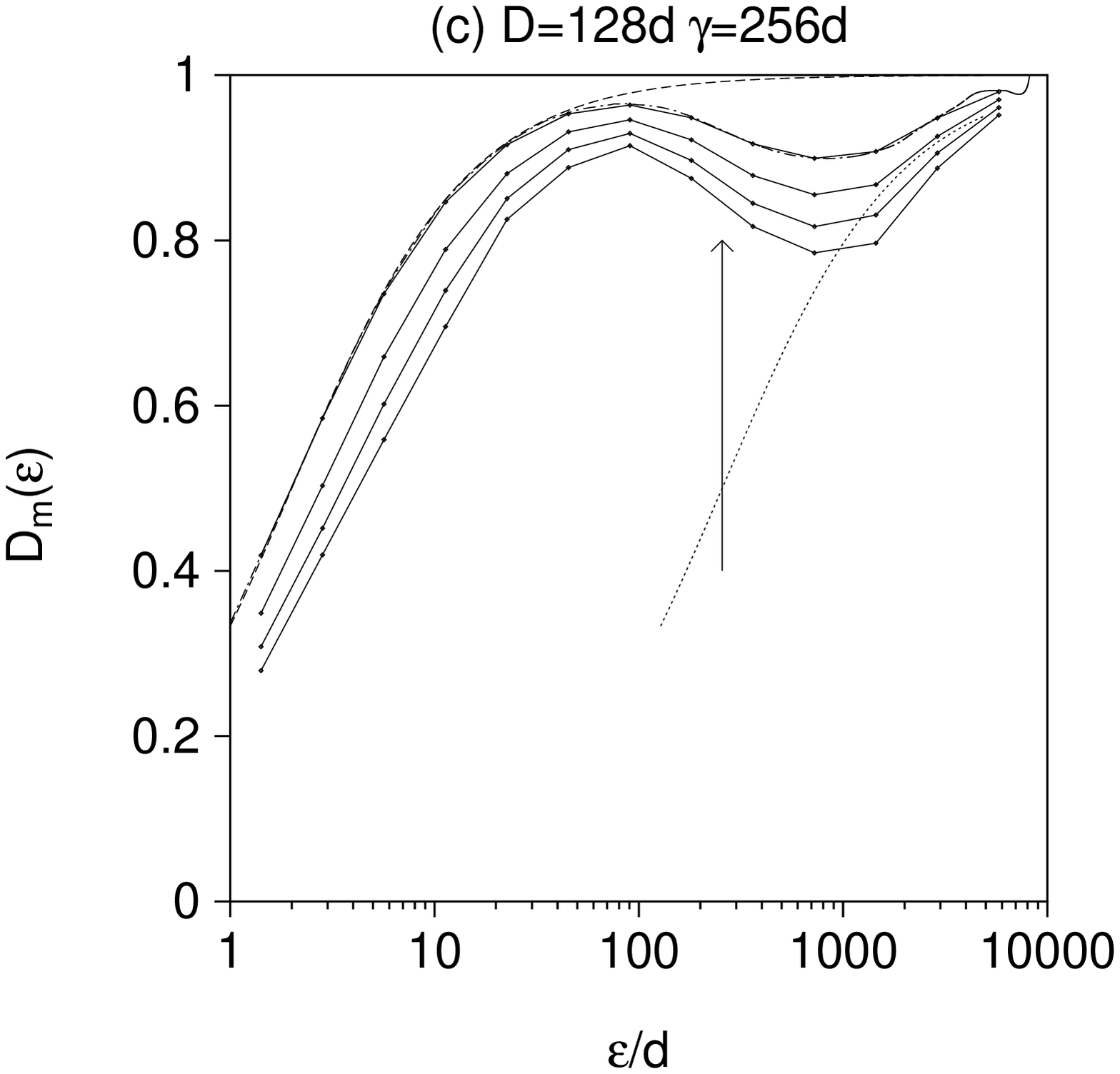,width=8cm}
  \end{minipage}
\end{center}
\begin{center}
  \begin{minipage}{8cm}
       \psfig{file=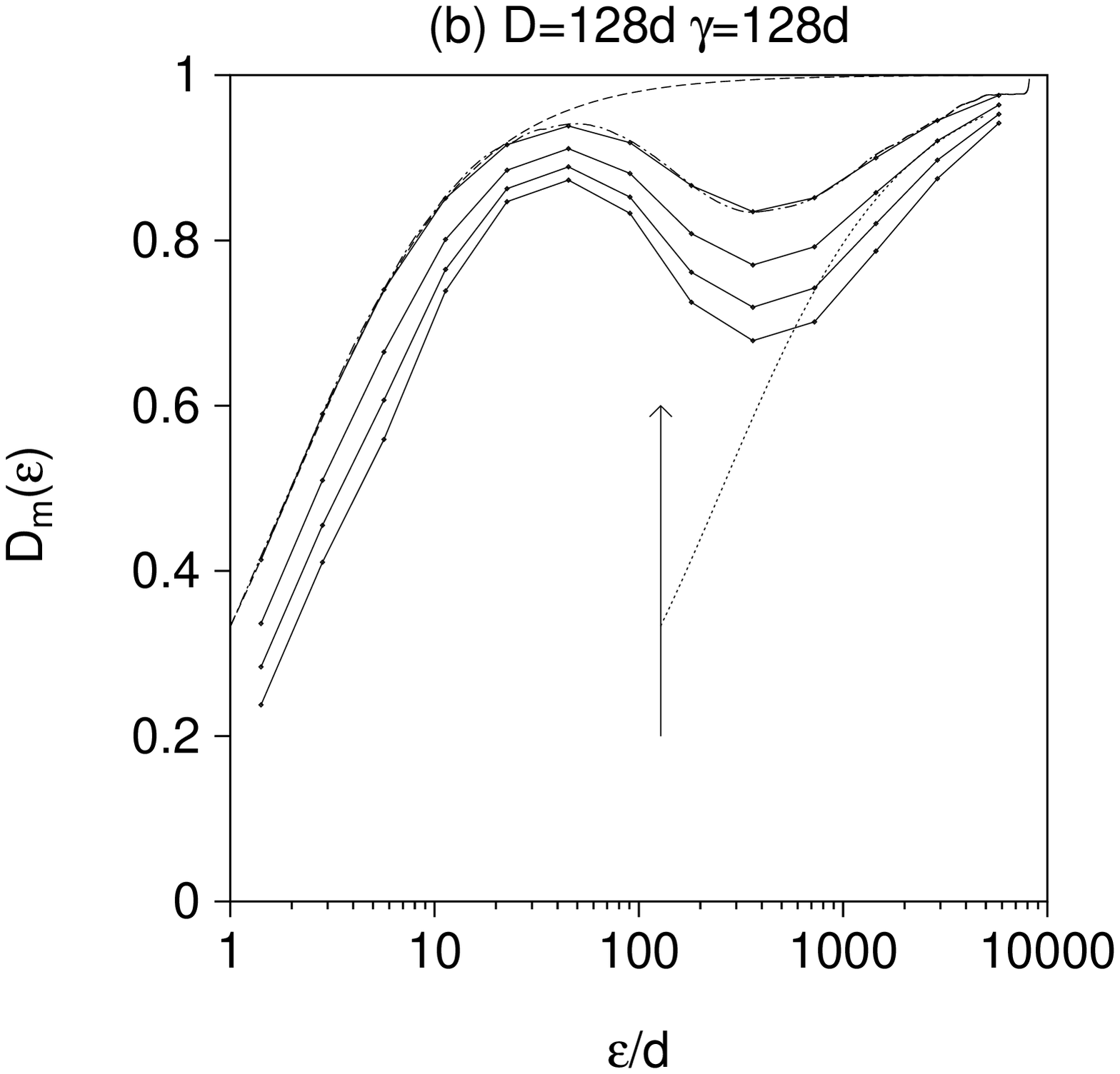,width=8cm}
  \end{minipage}
  \begin{minipage}{8cm}
       \psfig{file=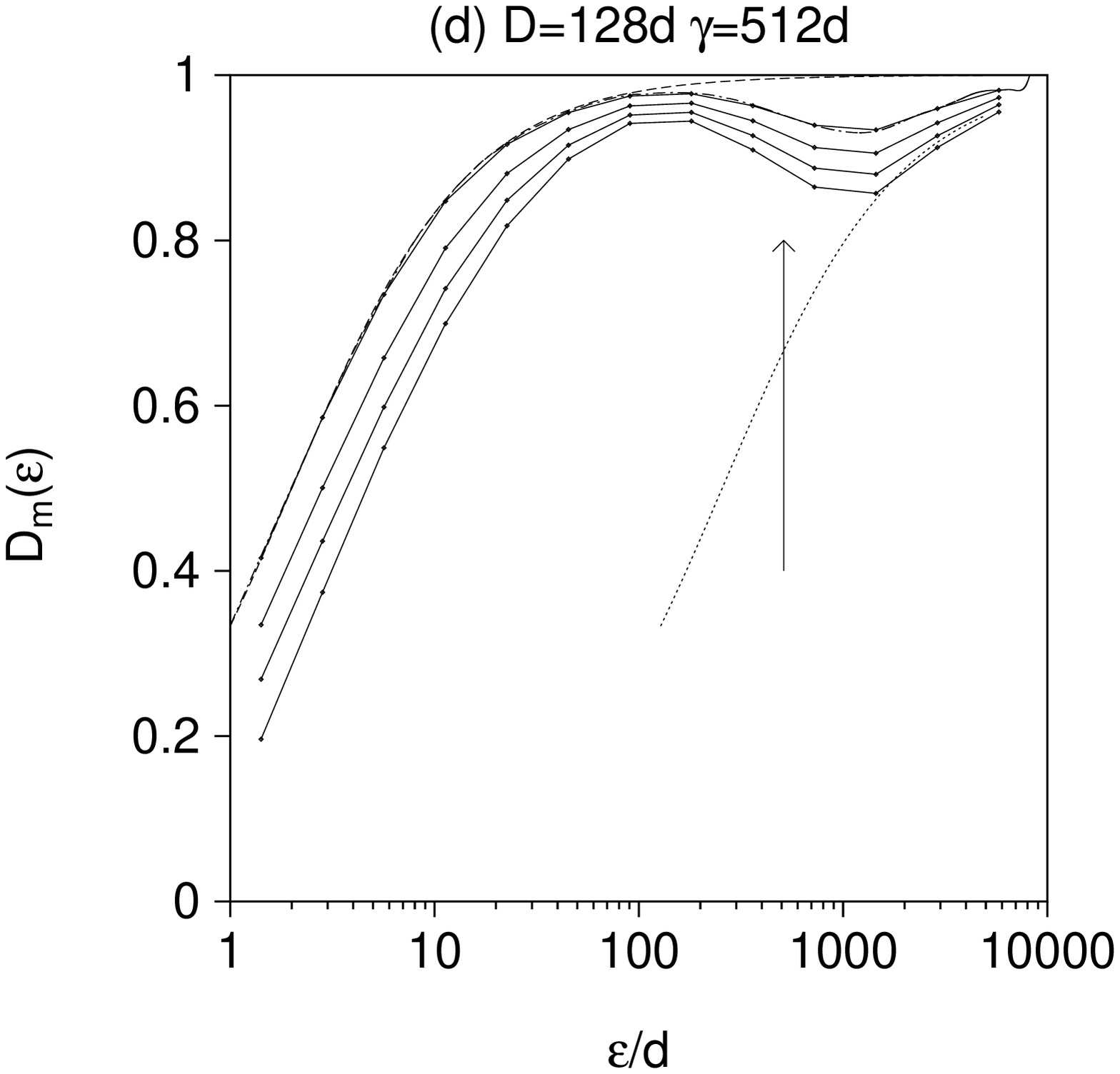,width=8cm}
  \end{minipage}
\end{center}
\caption{
The local scaling dimension $D_m(\epsilon)$ for the doorway damping
 model, corresponding to the partition functions plotted 
in Fig.\ \protect\ref{figpartition}.
The arrows  indicate the value of $\gamma$, which is chosen
as (a) $\gamma=64d$, (b) $128d$, (c) $256d$, and (d) $512d$.
The dashed curve represents $D_2(\epsilon)$
for the GOE.  The dotted curve represents the fluctuation associated
with the doorway coupling, which is taken the same as a GOE fluctuation 
with level spacing $D$ in the present model (see text). The
dashed-dotted curve is $D_2(\epsilon)$ evaluated by using the
derivative definition Eqs.\ (\protect\ref{scaledim}) 
and\ (\protect\ref{d2explicit}) instead of
the finite difference. 
}
\label{figmodelfra}
\end{figure}

Figures\ \ref{figpartition} and\ \ref{figmodelfra}   display
the partition functions $\chi_m(\epsilon)$ and the
local scaling dimensions $D_m(\epsilon)$, respectively, 
for the strength functions calculated with four different values 
of $\gamma$ and fixed value of $D=128d$. 
From the partition function
$\chi_m(\epsilon)$, one sees that the
fluctuation does not follow the power scaling law 
(or linear relation in the log-log plot). 
Correspondingly, the local scaling dimension 
$D_m(\epsilon)$ varies as a function of
$\epsilon$. 
The local scaling dimensions in the four cases  show
common features. Namely,
at small values of $\epsilon$,  $D_m(\epsilon)$ monotonically increases
with $\epsilon$. As $\epsilon$ increases further, $D_m(\epsilon)$ starts
to decrease, but it turns to increase again. At the largest value of
$\epsilon$ comparable to the whole spectrum range, $D_m(\epsilon)$
converges to the uniform limit $D_m(\epsilon)=1$.

Comparing with the local scaling dimension $D_m(\epsilon)$ for the
GOE strength fluctuation indicated with dashed curve in the figure, 
one sees that the behavior of $D_m(\epsilon)$ 
at small $\epsilon$ follows the GOE limit whereas
it deviates from the GOE
limit as $\epsilon$ increases. 
It is seen that 
the energy scale $\epsilon$ where 
 $D_m(\epsilon)$ starts to deviate from the GOE limit, 
which we denote $\epsilon^{*}$,  has strong
correlation to the value of  $\gamma$. 
As we discussed in terms of the level spacing distribution,
the GOE limit is realized as far as the small energy interval is
concerned. However, mixing between the background states (including the 
doorway states) takes place only within a finite energy interval,
that is, the spreading width $\gamma$ of the background states.
Therefore, the fluctuations far below
the energy scale $\gamma$ will obey the GOE limit, while
those comparable to or larger than $\gamma$ may deviate from the GOE limit.
In fact, one observes in Fig.\ \ref{figmodelfra}  that the energy scale 
$\epsilon^{*}$
where 
 $D_m(\epsilon)$ starts to deviate from the GOE limit 
is related approximately to $\gamma$ as $\epsilon^{*} \sim 1/5 \gamma$.

Figure\ \ref{figmodelfra} also shows clearly that 
the decrease of $D_m(\epsilon)$ at the intermediate energy scale
$\epsilon$ depends strongly on the parameter $\gamma$.
In fact, the value of $\epsilon$, denoted $\epsilon^{**}$,
where $D_m(\epsilon)$ decreases most sharply
corresponds well to the value of $\gamma$, indicated by
the arrow in the figure, and it moves as one changes $\gamma$.
Namely $\epsilon^{**} \sim \gamma$ is observed for all panels in the figure.
The close connection between  $\gamma$ and the decrease of
$D_m(\epsilon)$ can be explained as follows.
The presence of the doorway states causes the fine structure or the
clustering of the strength distribution.
When the spreading width $\gamma$ of the doorway states is small,
the doorway states form many peaks (clusters) where width of individual
peaks is given by $\gamma$. Using the bin width $\epsilon$ comparable
with the spreading width $\gamma$, the scaling analysis uncovers
the profile of the peaks associated with the doorways. Since
the profile will look the Lorentzian shape of width $\gamma$, 
the local scaling dimension $D_m(\epsilon)$ decreases sharply
around $\epsilon \sim \gamma$ as we found for the smooth Lorentzian
distribution (Sec.\ref{sec:lorentzian}). When $\gamma$ is larger
than the spacing $D$ of the doorway states, the peaks overlap with
each other and one cannot trace the individual doorway states.
However, the fluctuations of the strength distribution still
reflects the spreading width $\gamma$ of doorway states since
the fluctuations associated with the doorway states are  not
completely smeared out. Thus correspondence between
the spreading width $\gamma$ of doorway states and the point 
of steepest decrease in $D_m(\epsilon)$ holds also for the cases
of $\gamma \gtrsim D$. This situation corresponds to Figs.\
\ref{figmodelfra}(c) and (d).

The increase of $D_m(\epsilon)$ at the largest energy scales
arises from another features of the doorway states, that is, the
fluctuation in the collective-doorway coupling.
If we consider only the doorway states and neglect their coupling
to other backgrounds, the strengths are concentrated only on 
the doorway states, as shown in Fig.\ \ref{figvarstr}(b).  In this limit
the strengths of the doorways follow a GOE strength fluctuation, since the
collective-doorway coupling matrix elements is chosen as Gaussian random
variables. It should be noted that the energy scale
associated with this GOE is not the
level spacing $d$ of the whole spectrum, but  the spacing 
$D$ between the doorways, which is 
much larger than $d$. The dotted curve in Fig.\ \ref{figmodelfra} 
represents the
GOE strength fluctuation characterized by the level
spacing $D$ of the doorway states.
When the bin width 
$\epsilon$ is taken larger than the spreading width
$\gamma$ of the doorway states, the fluctuations at smaller energy scale
are smeared out, and only the fluctuation of the collective-doorway 
coupling remains.
In all cases in Fig.\ \ref{figmodelfra}, the local scaling dimensions
follow this curve at the largest energy scales, indicating that 
the strength fluctuation in this domain reflects the structure of
the collective-doorway coupling.

One may also notice that the dip of $D_m(\epsilon)$ 
around $\epsilon \sim \gamma$ becomes small 
and $D_m(\epsilon)$ approaches to the GOE limit as one increase the
value of $\gamma$. 
It is also noted that for small value of 
$\gamma$,  the local scaling dimension $D_m(\epsilon)$ 
for the higher moments $m$ takes significantly smaller value
around the intermediate $\epsilon$ 
than $D_m(\epsilon)$ for the lower moments (eg. $D_5$ vs. $D_2$),
while, for large $\gamma$, $D_m(\epsilon)$ for all moments
takes similar value. The higher moments are sensitive
to components  of the large strength $p_n$. Therefore,
small value of $D_m(\epsilon)$ for higher moments implies that the
strength distribution is very spiky. In other words
the strengths are clustered and form large peaks.
On the contrary, similar values
of $D_m(\epsilon)$ close to 1, obtained for large $\gamma$ values, implies
relatively uniform distribution of strengths.
The above two features characterize the way how the fine
structures (or clustering) associated with the doorway states
diminishes as $\gamma$ increases.

\begin{figure}
\begin{center}
  \begin{minipage}{8cm}
       \psfig{file=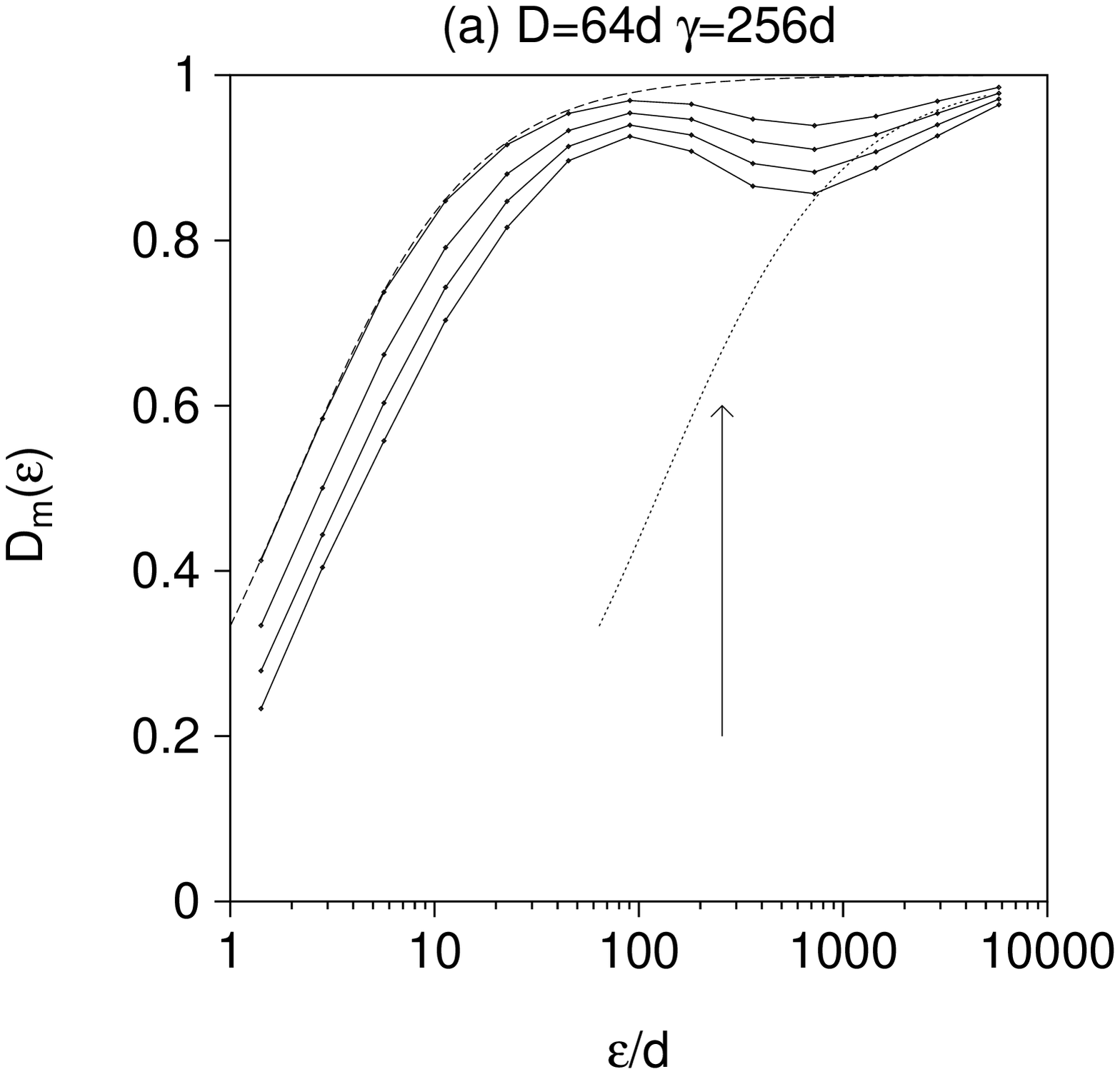,width=8cm}
  \end{minipage}
  \begin{minipage}{8cm}
       \psfig{file=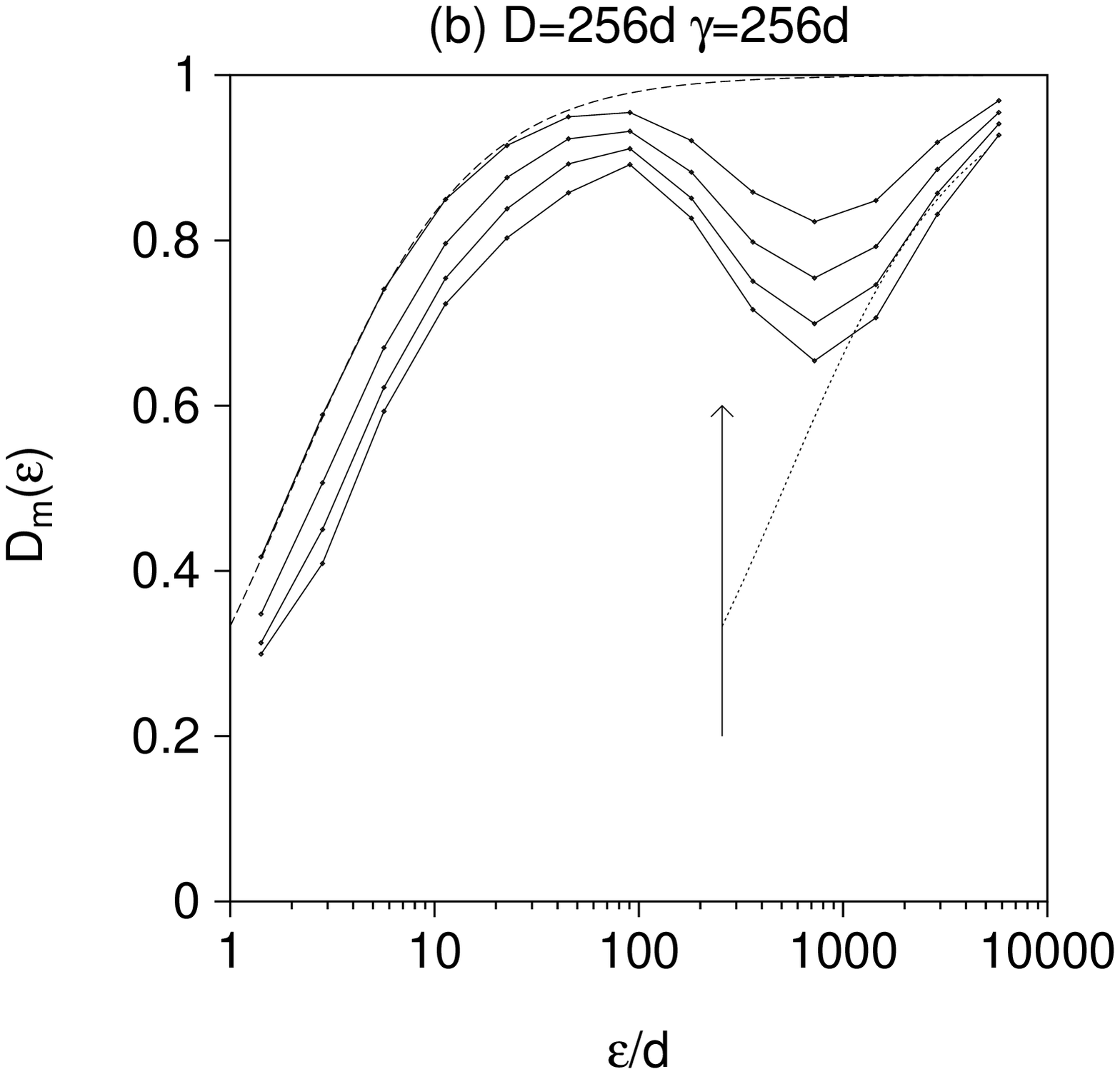,width=8cm}
  \end{minipage}
\end{center}
\caption{
The same as Fig.\ \protect\ref{figmodelfra} but 
for (a)$D=64d$ and $\gamma=256d$, and (b)$D=256d$ and $\gamma=256d$. 
}
\label{figotherdfra}
\end{figure}

In Fig.\ \ref{figotherdfra}, we show dependence on the
 parameter $D$, the spacing of the doorway states, with
the spreading width $\gamma$ being kept constant. It is seen that
the values of local scaling
dimension  $D_m(\epsilon)$ at larger energy scale $\epsilon$ depends on 
the spacing $D$ of doorway states.
When $D$ is small (e.g. Fig.\ \ref{figotherdfra}(a), where $D=64d$), 
$D_m(\epsilon)$
takes larger value close to the GOE limit than the case of larger $D$.
This implies that 
when the strength of 
each doorway states overlaps significantly with each other, namely, the
spacing $D$ of doorway states is much smaller than spreading width
$\gamma$, the fluctuation associated with the doorways tends to be
smeared. On the other hand, when $\gamma \lesssim D$, the fluctuation
associated with the doorways stands up clearly, then 
 $D_m(\epsilon)$ exhibits large deviation from the GOE limit.
However, it is more important to note that the spacing $D$ of the doorways
does not affect very much the behavior of  $D_m(\epsilon)$ at
smaller energy scale. In both cases of Fig.\ \ref{figotherdfra}, 
the values $\epsilon^{*}$ and $\epsilon^{**}$ 
where
$D_m(\epsilon)$ deviates from the GOE and decreases most sharply,
respectively, have relations to $\gamma$,
$\epsilon^{*}\sim 1/5\gamma$ and $\epsilon^{**}\sim\gamma$,
as was observed in Fig.\ref{figmodelfra}, indicating that 
these features hold in spite of the change of $D$ values.

In Fig.\ \ref{figmodelfra} we also plot with the dashed-dotted 
curve  $D_2(\epsilon)$ 
which is evaluated by means of the exact expression 
Eq.\ (\ref{d2explicit}) based on the derivative definition
Eq.\ (\ref{scaledim}). It is seen that 
difference between the exact evaluation
and the evaluation Eq.\ (\ref{approscaledim}) 
based on the finite difference is almost
indistinguishable. This justifies the use of the finite difference
evaluation of the local scaling dimension.

\subsubsection{$D_m(\epsilon)$ as a measure of doorway spreading width}

The close relation between the spreading width $\gamma$ 
and the profile of the local scaling dimension may be used as a tool
to estimate the
spreading width $\gamma$ of the doorway states.
The number of doorway states which are involved in the
damped collective states such as the giant resonances is not small.
In such cases, it is difficult to evaluate the spreading width
by distinguishing individual doorway states from the whole 
strength distribution since 
the strengths associated with each doorway
states may overlap with each other. On the other hand,
since the scaling analysis discussed in this paper does not
assume separation of the peaks, it is applicable in 
such overlapping cases. Indeed, the numerical analysis
presented above indicates that the characteristic behavior of
$D_m(\epsilon)$, namely the decrease 
around the energy scale comparable with $\gamma$, shows up
even in cases where the spreading width $\gamma$ is larger
than the spacing $D$ between the doorway states. Thus one may
estimate the spreading width $\gamma$ by locating the point 
$\epsilon^{**}$
where $D_m(\epsilon)$ decreases most steeply.

\begin{figure}
\begin{center}
  \begin{minipage}{8cm}
       \psfig{file=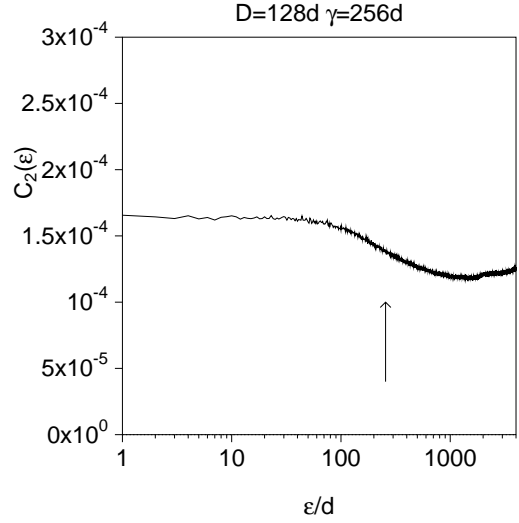,width=8cm}
  \end{minipage}
\end{center}
\caption{
The autocorrelation function $C_2(\epsilon)$ of the strength
functions for the doorway damping
model with $D=128d$ and $\gamma=256d$, corresponding to 
Fig.\ \protect\ref{figmodelfra}(c). The ensemble average is performed.
The arrow indicates the value of $\gamma$.
}
\label{figc2}
\end{figure}

It is instructive to compare  with the autocorrelation analysis,
which is often used to analyze the strength fluctuations.
As discussed in the previous section, there is the relation between
the autocorrelation analysis and the scaling analysis, in particular,
between the autocorrelation function $C_2(\epsilon)$ and the
local scaling dimension $D_2(\epsilon)$ for the second moment.
Figure\ \ref{figc2} shows the autocorrelation function $C_2(\epsilon)$
for $D=128d$ and $\gamma=256d$, which is also calculated for the
normalized strength distribution (e.g. Fig.\ \ref{figvarstr}(c)) and
averaged
over 60 realizations as done for the scaling analysis.
The autocorrelation function $C_2(\epsilon)$ exhibits the following
property. It has a broad peak centered around $\epsilon = 0$ which
is built upon a flat `background'. The broad peak contains the
information of the fine structure of the strength fluctuations
which arises from the doorway states. The width of the broad peak
corresponds well to the spreading width $\gamma$ of the doorway
states. It is helpful to remind that the autocorrelation function 
used to analyze the Ericson fluctuation in cross sections of
highly excited compound nuclei, where resonance overlaps with
neighboring ones. In that case, the width of the
autocorrelation function is related to the particle decay width
of individual states. In our case, on the other hand, 
the width of the autocorrelation function is related to the spreading width
of the doorway states. 
Comparing Fig.\ \ref{figmodelfra} and 
Fig. \ref{figc2}, one sees close relationship
between the local scaling dimension $D_2(\epsilon)$ and the
autocorrelation function $C_2(\epsilon)$, both of which decrease
sharply at the energy scale corresponding to $\gamma$.

\subsubsection{Analysis without ensemble average}

\begin{figure}
\begin{center}
  \begin{minipage}{8cm}
       \psfig{file=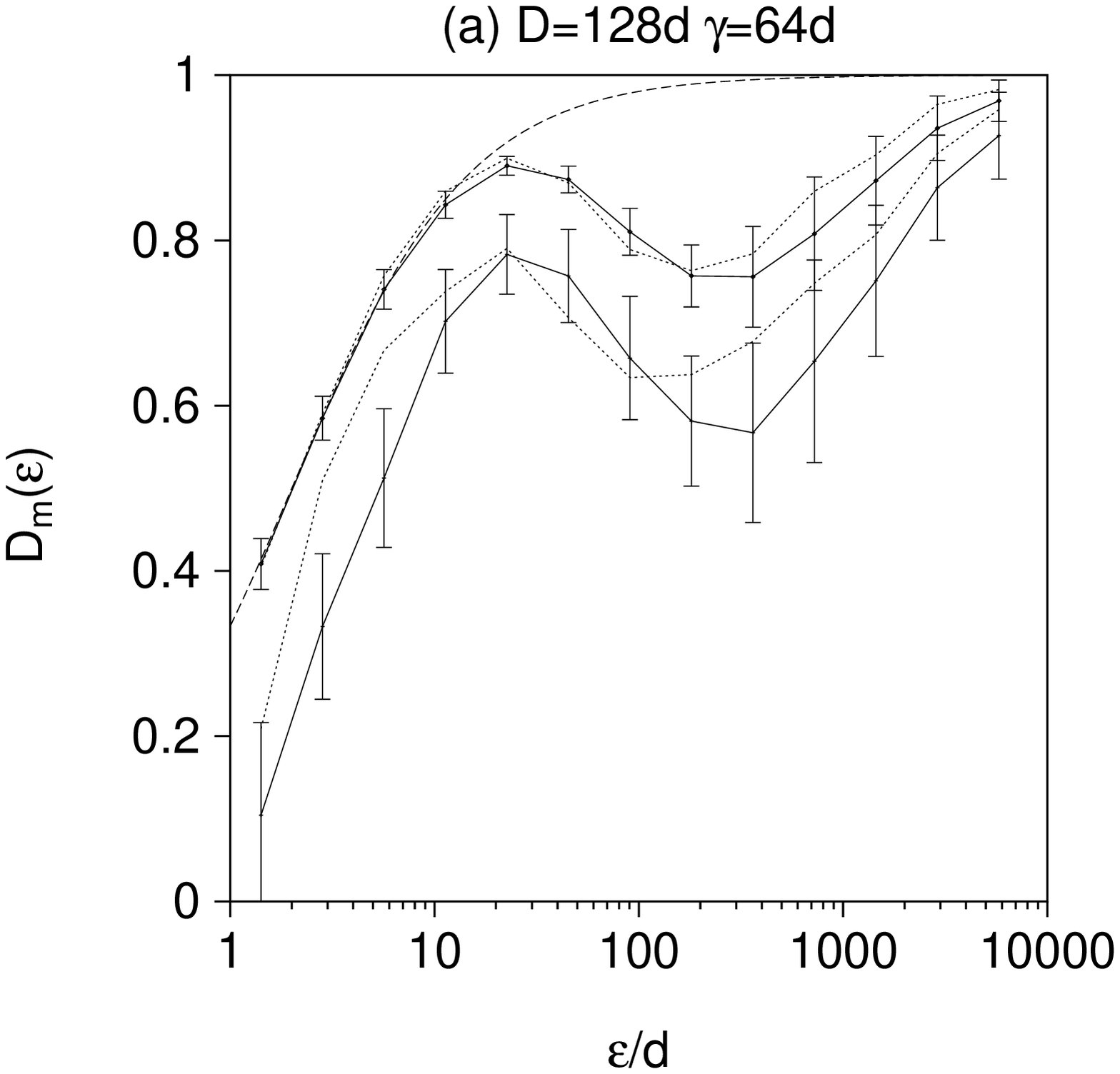,width=8cm}
  \end{minipage}
  \begin{minipage}{8cm}
       \psfig{file=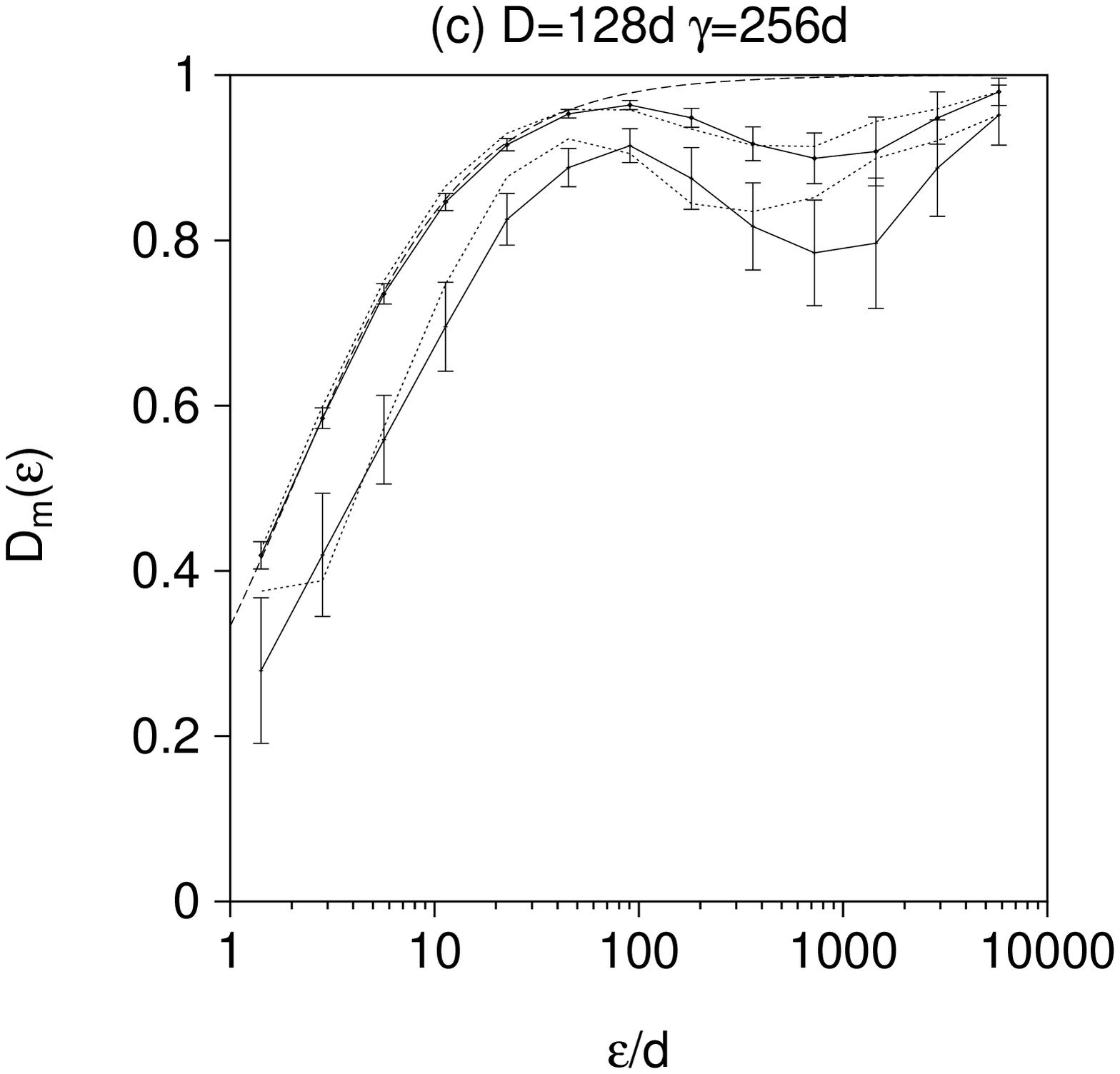,width=8cm}
  \end{minipage}
\end{center}
\begin{center}
  \begin{minipage}{8cm}
       \psfig{file=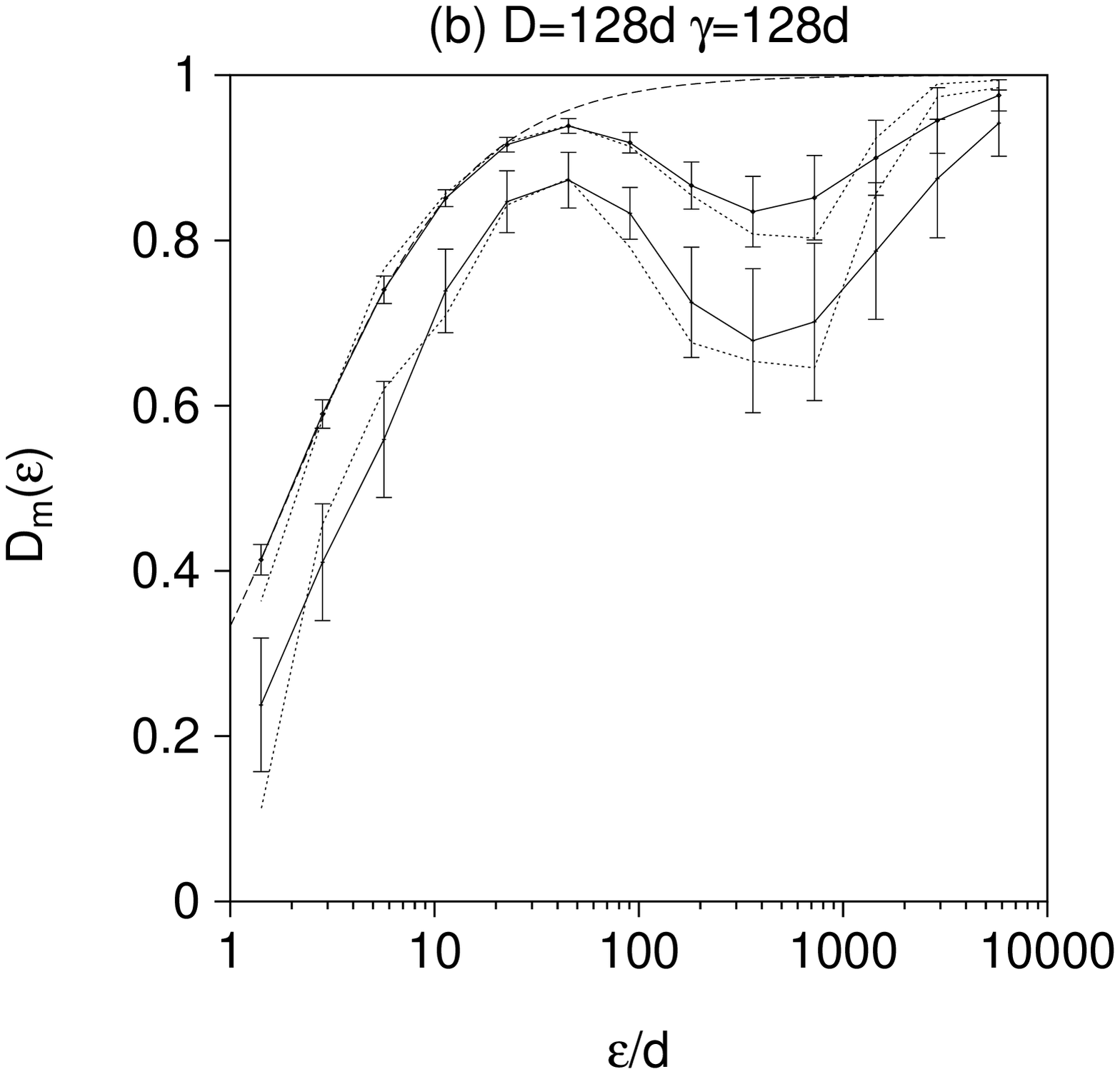,width=8cm}
  \end{minipage}
  \begin{minipage}{8cm}
       \psfig{file=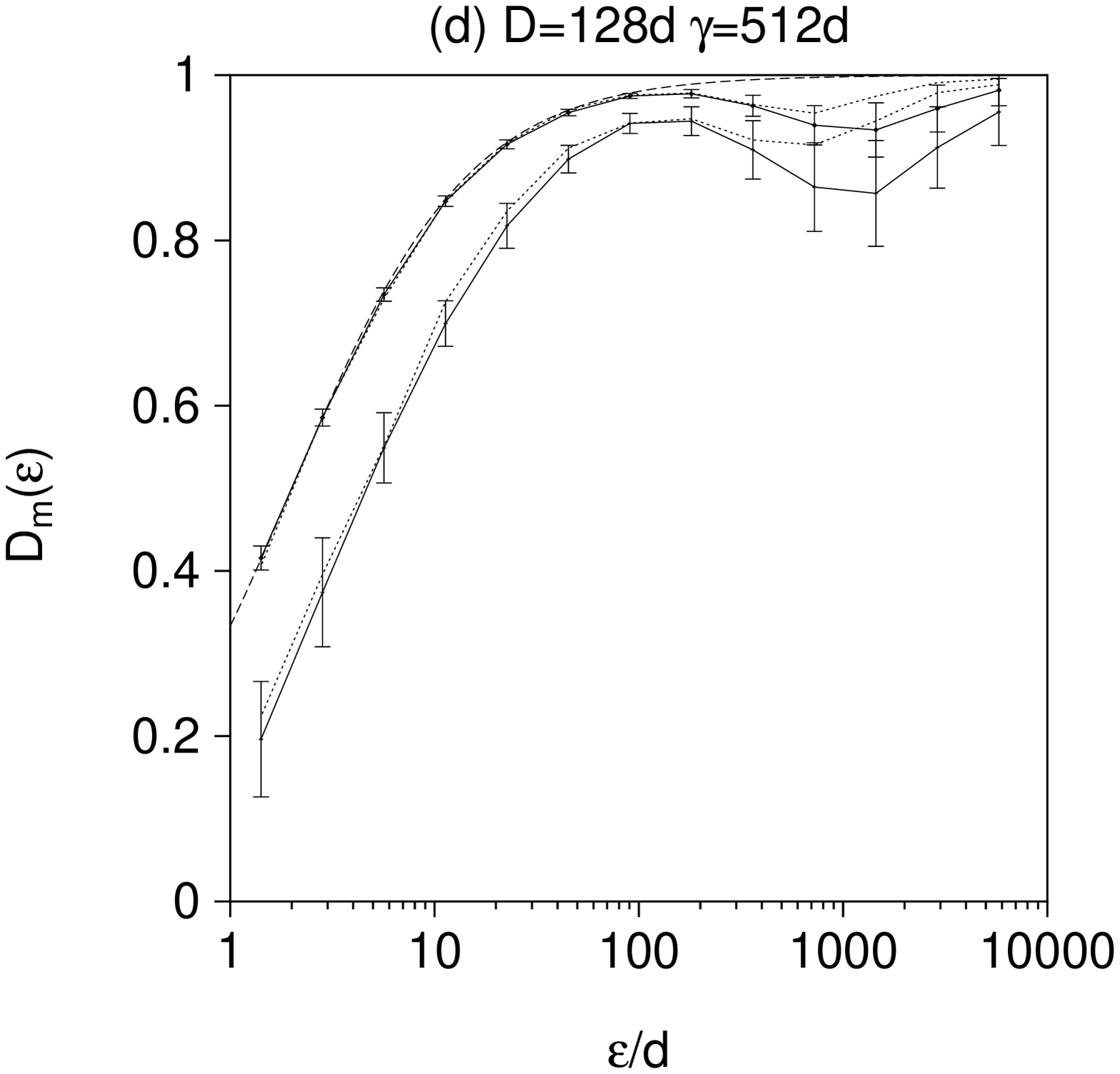,width=8cm}
  \end{minipage}
\end{center}
\caption{
Comparison of the local scaling dimensions $D_2(\epsilon)$ and 
$D_5(\epsilon)$ between the results obtained with the ensemble
average (solid curve) and those for a single realization of
spectra (dotted curve), for $D=128d$ and
(a) $\gamma=64d$, (b) $128d$, (c) $256d$, and (d) $512d$.
The bars indicate the standard deviation 
of  $D_2(\epsilon)$ and $D_5(\epsilon)$
associated with different realizations, which is evaluated by using the
ensemble of the 60 spectra.
}
\label{figflucd2}
\end{figure}

The results presented above are obtained by averaging over sixty
different strength functions which are calculated for every
realization of the random Hamiltonian. 
This is possible for the model systems. However, there will be a
situation where one has only a single experimental data for a given
nucleus so that one cannot perform ensemble average.
When one applies the scaling analysis to a single given
spectrum, the result will contain both the property 
which is common for the ensemble which the data belongs to and 
the behavior which is specific to the given realization of spectra. 
The latter can be regarded as fluctuation associated with 
different realizations. The fluctuation indicates an uncertainty
when we infer the ensemble averaged properties from analysis 
of a single spectrum.
To evaluate this uncertainty 
in the local scaling dimension $D_m(\epsilon)$, 
we calculate the local scaling dimensions $D_m(\epsilon)$ 
for each of 60 realizations of the model Hamiltonian, and
evaluate the average and the standard deviation in $D_m(\epsilon)$.
The results are plotted in Fig.\ \ref{figflucd2}. We also plot in 
the same figure an example of  $D_m(\epsilon)$ 
for an arbitrarily chosen realization.

It is seen that the local scaling dimension calculated for a single
realization shows  $\epsilon$ dependence which is qualitatively same
as the ensemble averaged $D_m(\epsilon)$. The standard deviation from 
the average is smaller than the typical size of  $\epsilon$ dependence of
$D_m(\epsilon)$ for the cases shown in the figure. 
It is seen that the deviation increases as $m$ increases, namely,
as the higher moments are concerned. 
The standard deviation in $D_m(\epsilon)$ is estimated approximately as
$\sigma(D_m) \approx \sqrt{3}m/\sqrt{N_{\rm tot}}$ 
by assuming independent Porter-Thomas fluctuation for the individual 
strengths $S_i$. 
Thus, the local scaling dimensions $D_m(\epsilon)$ calculated for a single
data are not affected very much by the fluctuation due to the
realizations and keep information on the background mechanisms
such as the spreading width $\gamma$ of the doorway states as
far as a sufficient number of levels are included in the analysis and
the low order moments are concerned.

\section{Conclusions}
\label{sec:conclusion} 

We proposed a new method to analyze fluctuation properties of the strength
functions by means of the scaling analysis similar to the multifractal
analysis. In order to treat the non-scaling fluctuations, we introduced
the local scaling dimension $D_m(\epsilon)$ defined 
as a function of a scaling parameter $\epsilon$, which is the 
energy bin width subdividing the strength function. This quantity
reduces to the generalized fractal dimension if a self-similar
structure exists in the strength fluctuation. The local scaling
dimension also has a close connection with the autocorrelation
function of the strength function.

Strength functions associated with damped excited modes such as
the giant resonances may exhibit  various fluctuations and
fine structures arising from different mechanisms including
the doorway structures of damping and the Porter-Thomas or GOE strength
fluctuations associated with individual levels.
Employing a model which mimics the doorway damping of the
giant resonances, we applied the method to analyze the strength
fluctuations associated with the doorway structures.
The close relation between the profile of the local scaling dimension
and the fluctuations embedded in the strength function
is studied in detail. We showed that the behavior of 
the local scaling dimension $D_m(\epsilon)$ as a function of 
the binning energy scale $\epsilon$
manifests clearly both the GOE fluctuation at the small energy scale and
the properties associated with the doorway states. For example,
the binning energy scale $\epsilon^{**}$ where the local scaling dimension 
$D_m(\epsilon)$ decreases
sharply corresponds to the spreading width $\gamma$ of the doorway states.
This demonstrates that the local scaling dimension can 
be an useful measure
to analyze the fluctuations and fine structures  of the strength function 
associated with the damping modes in highly excited nuclei.

It is in principle possible to apply the scaling analysis discussed in the
paper to the experimental spectra and also to 
strength functions in realistic descriptions of the giant
resonances in order to study the mechanism 
responsible for the strength fluctuations, 
such as the spreading width of the doorway states. The method itself
is, on the other hand, quite general and not restricted to the
giant resonances, and applicable to any strength fluctuation phenomenon
in  nuclei as well as in the other fields of quantum physics.

\acknowledgments

The authors acknowledge helpful discussions with
Kenichi Matsuyanagi, Toru Suzuki, and Shoujirou Mizutori.
The numerical calculations are performed at the
Yukawa Institute Computer Facility.
The work is supported by the Grant-in-Aid for 
Scientific Research from the Ministry of Education, Science
and Culture (No. 10640267).

\appendix

\section{Explicit Expression of Partition Function and Local Scaling
Dimension}
\label{appa}

We derive here an explicit expression of $\chi_m(\epsilon)$ 
given in terms of the strength $S_i$ and the energy $E_i$ of
individual levels.

The quantity $p_n$ defined in Eq.\ (\ref{defp}) can be written as
\begin{equation}
p_n=\sum_i S_i\theta_n(E_i),
\label{pdef2}
\end{equation}
by using a step function
\begin{equation}
\theta_n(x)\equiv\left\{
\begin{array}{ll}
1 & \qquad \mbox{for $x$ in $n$-th bin}\\
0 & \qquad \mbox{for $x$ not in $n$-th bin.}
\end{array}\right.
\label{thetan}
\end{equation}
Then, the partition function is also rewritten as
\begin{equation}
\chi_m(\epsilon)=
\sum_{i_1,\ldots,i_m}S_{i_1}\cdots S_{i_m}
\langle\sum_n\theta_n(E_{i_1})\cdots\theta_n(E_{i_m})\rangle_{\rm bin},
\label{chitheta}
\end{equation}
where $\langle\cdots\rangle_{\rm bin}$ represents an average with respect
to 
the bin boundary as explained in Sec.\ \ref{sec:localscaling}.
The quantity $\sum_n\theta_n(E_{i_1})\cdots\theta_n(E_{i_m})$
takes the value of either 1 or 0.
Denoting the difference between
the largest value and the smallest value among $E_{i_1},\cdots,E_{i_m}$
by $D_{i_1\ldots i_m}$, the probability of the quantity 
being 1 is evaluated as
$(\epsilon-D_{i_1\ldots i_m})/\epsilon$. In addition,  the quantity 
$\langle\cdots\rangle_{\rm bin}$ should be
 zero if  $D_{i_1\ldots i_m}$ is larger
than the bin width $\epsilon$. Thus, 
$\chi_m(\epsilon)$ can be calculated as
\begin{equation}
\chi_m(\epsilon)=
\sum_{i_1,\ldots,i_m}S_{i_1}\cdots S_{i_m}
\theta(\epsilon-D_{i_1\ldots i_m})
\frac{\epsilon-D_{i_1\ldots i_m}}{\epsilon},
\label{inichitheta2}
\end{equation}
\begin{equation}
\theta(x)\equiv\left\{
\begin{array}{ll}
1 & \qquad x > 0\\
0 & \qquad x \leq 0.
\end{array}\right.
\label{theta}
\end{equation}
As a result, the local scaling dimension $D_m(\epsilon)$ can be explicitly
written
as
\begin{equation}
D_m(\epsilon)=\frac{\sum_{i_1,\ldots,i_m}S_{i_1}\cdots S_{i_m}
\theta(\epsilon-D_{i_1\ldots i_m})D_{i_1\ldots i_m}
}{\sum_{i_1,\ldots,i_m}S_{i_1}\cdots S_{i_m}\theta(\epsilon-D_{i_1\ldots
i_m})
(\epsilon-D_{i_1\ldots i_m})}.
\label{dmexplicit}
\end{equation}
In particular, for the case of $m=2$, Eq.\ (\ref{inichitheta2}) and 
Eq.\ (\ref{dmexplicit}) lead to 
\begin{equation}
\chi_2(\epsilon)=
\sum_{ij}S_i S_j\theta(\epsilon -|E_i-E_j|)
\frac{\epsilon-|E_i-E_j|}{\epsilon},
\label{inichi2theta}
\end{equation}
and
\begin{equation}
D_2(\epsilon)=\frac{\sum_{ij}S_i S_j\theta(\epsilon-|E_i-E_j|)|E_i
-E_j|}{\sum_{ij}S_i S_j\theta(\epsilon-|E_i-E_j|)(\epsilon-|E_i
-E_j|)},
\label{d2explicit}
\end{equation}
respectively.

\section{Local Scaling Dimensions for GOE Strength Fluctuations}
\label{appb}

The local scaling dimension associated with the GOE strength fluctuation 
can be calculated as follows under the assumption that the energy levels
are
equally spaced $E_i=id~(i=1,2,..., N_{\rm tot})$.
Since the strength $S_i$ of
$i$-th level is written as $S_i=x_i^2$ in terms of amplitudes $x_i$
which obeys the GOE amplitude distribution 
with dimension $N_{\rm tot}$\ \cite{brody}
\begin{equation}
 P_{\rm GOE}(x_1,...,x_{N_{\rm tot}})=\pi ^{-{N_{\rm tot} \over 2}}\Gamma(
 {N_{\rm tot} \over 2})\delta(1-\sum_{i=1}^{N_{\rm tot}}x_i^2),
\label{distgoe}
\end{equation}
the probability distribution
of \{$p_n$\} defined by Eq.\ (\ref{defp}) can be written as
\begin{equation}
f(p_1,...,p_L)=\int ...\int dx_1...dx_{N_{\rm tot}}
\delta(p_1-\sum_{i\in{\rm 1st~bin}}x_i^2)...
\delta(p_L-\sum_{i\in L{\rm -th~bin}}x_i^2)P_{\rm GOE}(x_1,...,x_{N_{\rm
tot}}).
\label{probpat}
\end{equation}
Considering the case where the bin width $\epsilon=ld$ ($l$ being 
 even integer), that is,
each bin contains $l$ levels, and using the variable $l$ as the scaling
parameter instead of $\epsilon$, 
the ensemble average of the partition function can be calculated
as
\begin{eqnarray}
\overline{\chi_m(l)} &=& \int\cdots\int dp_1\cdots dp_L
\sum_{n=1}^L p_n^m f(p_1,\cdots,p_L)\nonumber\\
&=&{N_{\rm tot} \over l}{\Gamma({N_{\rm tot} \over 2})
\Gamma({1\over 2}(2m+l))\over\Gamma({l \over 2})\Gamma({1\over 2}
(N_{\rm tot}+2m))}. 
\label{patgoe}
\end{eqnarray}
Thus, the local scaling dimension derived from Eq.\ (\ref{approscaledim})
reads
\begin{equation}
D_m(\sqrt{2}l)={1\over (m-1)\log 2}(-\log 2+\sum_{i=l}^{m+l-1}\log i-
\sum_{i=l/2}^{m+l/2-1}\log i).
\label{dimgoedif}
\end{equation}
If we use Eq.\ (\ref{scaledim}) by taking $l$ as a continuous variable,
 the local scaling
dimension can be written as
\begin{equation} 
D_m(l)={1\over m-1}\sum_{i=1}^{m-1}{l\over l+2i}.
\label{dimgoeder}
\end{equation}
As is seen from Eqs.\ (\ref{dimgoedif}) and (\ref{dimgoeder}),
the local scaling dimension for the  GOE strength fluctuation is
independent of
the matrix dimension. In deriving the above results,
we did not perform an average with respect to the bin boundary for the
partition
function. This is not necessary since the ensemble average has the
same effect for the case of GOE.

In the above evaluation we neglected the fluctuation of the energy
levels. The analytic expression which includes both the strength and
the energy level fluctuations can be derived for the local scaling 
dimension of the second moment, 
$D_2(\epsilon)$.

Using Eq.\ (\ref{inichi2theta}), we write the partition function as
\begin{equation}
\epsilon\chi_2(\epsilon)=\int d\epsilon'\int d\epsilon''
\theta(\epsilon-|\epsilon'-\epsilon''|)(\epsilon-|\epsilon'-\epsilon''|)
A(\epsilon',\epsilon''),
\label{epsichi2}
\end{equation}
where the function $A(\epsilon',\epsilon'')$ is defined by
\begin{equation}
A(\epsilon',\epsilon'')\equiv\sum_{ij}S_iS_j\delta(\epsilon'-E_i)
\delta(\epsilon''-E_j).
\label{defa}
\end{equation}
Now, we consider the ensemble average of the function
$A(\epsilon',\epsilon'')$.
Since strengths and energies can be averaged independently, and
the average of $S_i^2$ and that of $S_iS_j$ ($i\ne j$) do not depend on
$i$ and $j$, the result of ensemble average is given by
\begin{equation}
\overline{A(\epsilon',\epsilon'')}=\overline{S_i^2}\delta(\epsilon'
-\epsilon'')\overline{\sum_i\delta(\epsilon'-E_i)}+
\overline{S_iS_j}\overline{\sum_{i\ne j}\delta(\epsilon'-E_i)
\delta(\epsilon''-E_j)}.
\label{bara}
\end{equation}
Since the average level density is normalized ( unfolded), 
\begin{equation}
\overline{\sum_i\delta(\epsilon'-E_i)}=1,
\label{averho}
\end{equation}
and using the Dyson's two-point cluster function \cite{brody,bohigas},
\begin{equation}
\overline{\sum_{i\ne j}\delta(\epsilon'-E_i)\delta(\epsilon''-E_j)}=
1-Y(\epsilon'-\epsilon''),
\label{bohigas}
\end{equation}
the ensemble average of Eq.\ (\ref{epsichi2}) can be written as,
\begin{equation}
\epsilon\overline{\chi_2(\epsilon)}/N_{\rm tot}=\overline{S_i^2}\epsilon+
\overline{S_iS_j}\bigl(\epsilon^2-\int_{-\epsilon}^\epsilon
dx(\epsilon-|x|)Y(x)\bigr).
\label{aveepsichi2}
\end{equation}

For the GOE with large $N_{\rm tot}$ limit, we have\ \cite{brody,bohigas}
\begin{equation}
\overline{S_i^2}=\alpha\overline{S_iS_j} \quad (\alpha=3),
\label{varstrgoe}
\end{equation}
and
\begin{equation}
Y(\epsilon)=S(\epsilon)^2-J(\epsilon)D(\epsilon),
\label{clustergoe}
\end{equation}
where the functions $S(\epsilon)$, $J(\epsilon)$, and 
$D(\epsilon)$ are given by
\begin{eqnarray}
S(\epsilon)&\equiv&{\sin(\pi\epsilon)\over \pi\epsilon},\\
\label{defs}
D(\epsilon)&\equiv&{dS(\epsilon)\over d\epsilon},\\
\label{defd}
J(\epsilon)&\equiv&\int_0^\epsilon d\epsilon'S(\epsilon')-{1\over 2}
{\rm sgn}(\epsilon).
\label{defj}
\end{eqnarray}
Using (\ref{aveepsichi2})-(\ref{defj}), the local scaling
dimension for the GOE strength fluctuation results in
\begin{equation}
D_2(\epsilon)={G(\epsilon)\over F(\epsilon)},
\label{d2goe}
\end{equation}
where
\begin{equation}
G(\epsilon)=\epsilon^2-{\sin(\pi\epsilon)\over \pi}(1+{2\over \pi}{\rm Si}
(\pi\epsilon))+{{\rm Si}(\pi\epsilon)\over \pi}(1-{{\rm Si}(\pi\epsilon)
\over \pi})+{4\over \pi^2}\int_0^{\pi\epsilon}\cos y{\rm Si}ydy,
\label{defG}
\end{equation}
\begin{equation}
F(\epsilon)=\epsilon(\epsilon+\alpha+1)+{{\rm Si}(\pi\epsilon)
\over \pi}({{\rm Si}(\pi\epsilon)\over \pi}+{4\over \pi}\sin(\pi\epsilon))
-{4\epsilon\over \pi}({\rm Si}(2\pi\epsilon)
-{\sin^2(\pi\epsilon)\over \pi\epsilon})-{4\over \pi^2}
\int_0^{\pi\epsilon}\cos y{\rm Si}ydy,
\label{defF}
\end{equation}
\begin{equation}
{\rm Si}x\equiv\int_0^x{\sin y \over y}dy.
\label{defsi}
\end{equation}

One sees that when $\epsilon$ is sufficiently larger than the
level spacing, or practically when $\epsilon \gtrsim 10$,  
Eq.\ (\ref{d2goe}) leads to
\begin{equation}
D_2(\epsilon)\simeq{\epsilon \over \epsilon+\alpha-1},
\label{d2bigepsi}
\end{equation}
which is the same as Eq.\ (\ref{dimgoeder}) with $m=2$.
This indicates that the fluctuation of energy levels does not affect
the local scaling dimension for $\epsilon$ which is larger than
ten times of level spacing. Note that the effect of the
energy level fluctuation on $D_2(\epsilon)$ is not very large also
for the region $1<\epsilon<10$. At $\epsilon =4$, Eq.\ (\ref{d2bigepsi})
gives
$D_2=0.67$ while inclusion of the level fluctuation (Eq.\ (\ref{d2goe}))
leads to
$D_2=0.62$.

Equation (\ref{aveepsichi2}) can be used also for the Poisson
fluctuation, for which the strength is constant
$ \overline{S_i^2}=\overline{S_iS_j}$,
and there is
no correlation among energy levels $Y(x)=0$. Then, 
Eq.(\ref{aveepsichi2}) leads to, up to a constant factor,
\begin{equation}
\overline{\chi_2(\epsilon)}=\epsilon+1,
\label{chi2poi}
\end{equation}
and the local scaling dimension
$D_2(\epsilon)$ for the second moment is given by
\begin{equation}
D_2(\epsilon)={\epsilon\over \epsilon+1},
\label{d2poi}
\end{equation}
which is equivalent to Eq.\ (\ref{dmpoisson}) with $m=2$.

\break

%\break

\break

\break

\break

\break

\break

%\break

\break

\end{document}